\documentclass[journal,twocolumn,10pt,twoside]{IEEEtran}

\IEEEoverridecommandlockouts
\usepackage{cite}
\usepackage{amsmath,amssymb,amsfonts}
\usepackage{algorithmic}
\usepackage{graphicx}
\usepackage{textcomp}
\usepackage{bbm}
\usepackage{nicefrac}
\usepackage{color}
\usepackage{hyperref}
\usepackage{caption}

\def\BibTeX{{\rm B\kern-.05em{\sc i\kern-.025em b}\kern-.08em
		T\kern-.1667em\lower.7ex\hbox{E}\kern-.125emX}}

\newtheorem{assumption}{Assumption}

\newtheorem{fact}{{Fact}}
\newtheorem{theorem}{{Theorem}}
\newtheorem{lemma}{{Lemma}}

\newcommand{\cM}{\mathcal{M}}
\newcommand{\cN}{\mathcal{N}}
\newcommand{\cL}{\mathcal{L}}
\newcommand{\cI}{\mathcal{I}}
\newcommand{\fm}{\mathfrak{m}}
\newcommand{\fu}{\mathfrak{u}}

\newcommand{\yy}{s} 

\title{\LARGE \bf
	Distributed Mechanism Design for Network Resource Allocation Problems
}

\author{Nasimeh Heydaribeni and Achilleas Anastasopoulos%
	\thanks{This work was supported in part by NSF Grant ECCS-1608361.}
	\thanks{This work has been partially presented in~\cite{HeAn18} and~\cite{heydaribeni2018distributed}.}%
	\thanks{The authors are with the Department of Electrical Engineering and Computer Science, University of Michigan, Ann Arbor, MI, 48105 USA {\tt\small {heydari,anastas}@umich.edu}}
}
\begin{document}
	
	\maketitle
	\thispagestyle{empty}
	\pagestyle{empty}

	\maketitle
	
	\begin{abstract}
		In the standard Mechanism Design framework, agents' messages are gathered at a central point and allocation/tax functions are calculated in a centralized manner, i.e., as functions of all network agents' messages.
		This requirement may cause communication and computation overhead and necessitates the design of mechanisms that alleviate this bottleneck.
		
		We consider a scenario where message transmission can only be performed locally so that the mechanism allocation/tax functions can be calculated in a decentralized manner.
		Each agent transmits messages to her local neighborhood, as defined by a given message-exchange network, and her allocation/tax functions are only functions of the available neighborhood messages.
		This scenario gives rise to a novel research problem that we call ``Distributed Mechanism Design".
		
		In this paper, we propose two distributed mechanisms for network utility maximization problems that involve private and public goods with competition and cooperation between agents. As a concrete example, we use the problems of rate allocation in networks with either unicast or multirate multicast transmission protocols.
		The proposed mechanism for each of the protocols fully implements the optimal allocation in Nash equilibria and its message space dimensionality scales linearly with respect to the number of agents in the network.
	\end{abstract}

	\section{Introduction}

	Allocation of scarce resources in networks has been a topic of intensive research in the last fifty years.
	This problem is often formulated as a network utility maximization (NUM) problem~\cite[Ch.~2]{srikant2013communication}
	where the designer is seeking the optimal allocation of resources that maximizes the social welfare.
	The complexity of this problem, especially for large networks of heterogeneous and strategic agents with privacy constraints
	stems from the fact that agents may not be willing to share some of their private information related to their utilities. Hence, appropriate incentives (taxes/subsidies) have to be put in place to incentivize agents to reveal some part of their private information relevant to the welfare optimization problem.
	A useful mathematical framework for this setting is mechanism design (MD)~\cite{HuRe06,BoKrSt15} that has been widely utilized in such areas of research as market allocations~\cite{groves,hurwicz1979outcome,YaHa05}, rate and resource allocations~\cite[Sec.~2.7]{srikant2013communication}\cite{basar,jain2010efficient,KaTe13,KaTe15,SiAn17b}, spectrum sharing~\cite{huang2006auction,wang08,wang10}, data security~\cite{khalili2017designing}, power allocation in wireless networks~\cite{vanderschaar07,demosshruti}, demand-side management in the power grid~\cite{fahrioglu2000,caron2010,samadi2012}, etc.

	In the standard MD framework (Hurwicz-Reiter~\cite{HuRe06}) agents are transmitting messages to a central authority. The central authority, upon receiving all these messages, determines allocation and taxes/subsidies for the agents of the network. Equivalently, all agents broadcast their messages to everyone else and then the allocation and taxes can be generated (and verified) by everyone in the network. Clearly, this arrangement results in a significant communication overhead due to message transmission of agents to the central authority or to each other, and this problem becomes more pronounced the larger the network.
	The motivation for this work is the more realistic scenario where such message transmission to a central authority (or equivalently, broadcasting of messages) cannot take place due to network communication constraints. To investigate this problem, we consider a setting in which agents are only allowed to transmit their messages to their neighboring agents. In this setting, neighborhoods are defined through an underlying message-exchange network.  Consequently, each agent can determine her allocation and tax based on the messages she hears locally and therefore, there is no need for a central authority to evaluate these functions.  This implies that, unlike standard mechanisms, the designed allocation and tax functions cannot have the whole message space as their domain; rather the allocation and tax function for each agent should only depend on the neighborhood messages. This additional restriction gives rise to a novel research direction that we call ``Distributed Mechanism Design" (DMD).

	A complementary view of DMD stems from the literature of distributed optimization (e.g.,  \cite{everett1963generalized,gabay1975dual,eckstein1992douglas,boyd2011,kraning2014dynamic,zhu2015distributed,dantzig2016linear}). In distributed optimization agents do exchange local messages in order to solve a centralized allocation problem. It is assumed however, that agents are not strategic--in fact they are automata--and follow a predefined message exchange algorithm.
	DMD can be thought of as the generalization of distributed optimization to account for strategic agents, i.e., for settings where we can no longer assume that agents will follow a distributed message passing algorithm unless the designer puts in place appropriate incentives for them to do so.

	In this paper we consider two network resource allocation problems with increasing degree of sophistication, formulated as NUM problems. Although the models presented are abstract, we present them through two concrete applications.
	In particular, we consider rate allocation for data transmission on a network which operates either with a unicast transmission protocol (UTP) or a multirate multicast transmission protocol (MMTP).
	For each of these protocols, a distributed mechanism is proposed for efficient rate allocation.
	The distributed mechanism proposed for MMTP is an extended and modified version of the distributed mechanism for UTP.
	For this reason, we present these two mechanisms side-by-side and highlight the main techniques used and the differences between them throughout the presentation.
	
	The contributions of this paper are as follows. Both of the proposed mechanisms are (a) distributed, (b) they fully implement the optimal allocation in Nash equilibria (NE) (i.e., there are no extraneous non-efficient equilibria in the induced game), and (c) their total message space dimension grows linearly with respect to the number of network agents. Furthermore, the mechanisms are (d) individually rational and (e) weak budget balanced at NE.

	\subsection{Relevant Literature}

	A non-distributed mechanism for efficient rate allocation with UTP has been proposed in~\cite{jain2010efficient,KaTe15,SiAn15b} and with MMTP in~\cite{KaTe13,SiAn17b}.
	In this paper the models we consider closely follow the models in these works but the mechanisms differ in a fundamental way since our focus is designing distributed mechanisms.
	The current work builds on distributed mechanisms for Walrasian and Lindahl allocation in private and public goods, respectively, that were proposed in~\cite{SiAn17d}\cite[Ch. 4]{Si17}.
	We have utilized an idea similar to the radial allocation~\cite{YaHa05,basar,SiAn17b} to achieve feasibility at NE.
	Unlike the mechanism in~\cite{SiAn17d} with message dimensionality per user that grows linearly with the number of users in the whole network, in this work, the message dimensionality of each agent grows linearly only with respect to the size of her neighborhood.

	There is a line of research in the computer science literature by the name distributed algorithmic mechanism design (DAMD)~\cite{feigenbaum2001sharing,feigenbaum2002distributed,sami2003distributed}\cite[Ch. 14]{nisan2007algorithmic}. We caution the reader not to confuse DAMD with DMD. The mechanisms in DAMD impose no restrictions on the domain of the allocation and tax functions. Indeed these functions can depend on the entire message from all users. The ``distributed'' aspect of DAMD pertains to the fact that an algorithm is designated to collect and disseminate all these messages to the users of the network so that they can all evaluate these complex functions. In DMD however, the allocation and tax functions are explicitly designed so that they only depend on messages from neighboring agents. Another difference is that in DAMD the message exchange network coincides with the network implied by the allocation problem (e.g., in~\cite{feigenbaum2001sharing} messages are exchanged between neighboring agents on the multicast tree) while in DMD, as will be evident in Section~\ref{section2}, the message exchange network can be arbitrary.

	A related line of work in distributed optimization attempts to resolve ``privacy'' issues by means of dithering (i.e., adding noise to) the exchanged messages or the objective function as in~\cite{HuMiVa15, nozari2016differentially, cortes2016differential, han2017differentially}.
	In these works, agents are given some privacy guarantees in that the distributed algorithm does not fully reveal their private information. However, the agents are still considered non-strategic automata, i.e., it is assumed that they follow the prescription of the algorithm even if there is an incentive to deviate.

	We conclude the discussion on the relevant literature by pointing out that the games induced by DMD fall under the class of ``graphical games''~\cite{kearns2001graphical,jackson2010social}. This property may have some consequences on the complexity of  off-line evaluation of the NE, which however, is not of importance in our work.

	The rest of the paper is structured as follows. In Section~\ref{section2}, the model and problem formulation for both of the network transmission protocols are discussed. Section~\ref{section3} presents both of the distributed mechanisms by characterizing the message-exchange network and defining messages, allocation and tax functions.
	A specific example is discussed in detail in Section~\ref{section_example} in order to clarify the general definitions of the distributed algorithms.
	In Section~\ref{section4}, the properties of the designed mechanisms are derived and the main results are presented. In Section~\ref{section5}, for each of the two mechanisms, an alternative mechanism is presented by relaxing an assumption on the message-exchange network. We conclude and comment on the message dimensions in Section~\ref{section6}.
	All of the proofs of intermediate lemmas can be found in the Appendix and also at an extended version of this paper~\cite{ARXIV-VERSION-J}.

	\section{Model} \label{section2}
	
	The two abstract centralized optimization problems for which distributed mechanisms will be developed in the following sections are precisely defined in~\eqref{CPu} and~\eqref{CPm0}, respectively.
	In order to make the discussion more relevant to resource allocation in data transmission networks, we now present two concrete applications that will serve as prototypes in the subsequent discussion.
	In particular, we consider a data transmission network with two different transmission protocols, Unicast Transmission Protocol (UTP) and Multirate Multicast Transmission Protocol (MMTP) and for each one of them an optimization problem for efficient data rate allocation is presented.
	We are following closely the models developed in
	~\cite[Sec. 2.7]{srikant2013communication}\cite{jain2010efficient,KaTe13,KaTe15,SiAn17b}. The network consists of multiple sources in the set $\mathcal{K}=\{1,...,K\}$ and strategic receivers in the set $\cN=\{1,...,N\}$, which will be referred to as agents.
	Each agent has one designated source from which it receives content, and each source can send content to multiple agents possibly with different data rates. The vector of allocated rates to all agents is denoted by $x=(x_1,...,x_N)\in \mathbb{R}_+^N$, where $x_i$ is the data rate of agent $i$.
	Based on its allocated rate, $x_i$, each agent receives a  private utility (satisfaction) $v_i(x_i)$.
	The following assumptions are imposed on the utility functions. For every $i \in \cN$, $v_i(\cdot)\in \mathcal{V}_0$, where $\mathcal{V}_0$ is the set of strictly concave, monotonically increasing, twice differentiable $\mathbb{R}_+\rightarrow \mathbb{R}$ functions with continuous second derivatives.  The valuation function $v_i(.)$ is the private information of agent $i$.
	The network links are denoted by $\cL=\{1,...,L\}$, each of which has capacity $c^l>0$. Agent $i$'s data stream is transmitted via links $\cL_i \subset \cL$ with $|\cL_i|=L_i$. The routing has been established in advance and it is considered fixed for the problem of interest in this paper. For each link $l$, agents using it are denoted by $\cN^l$ with $|\cN^l|=N^l$.
	The designer's goal is to maximize the social welfare which is the summation of agents' valuation functions. This is done by determining the efficient $x$ that is consistent with the capacity constraints of the network which arise from the specific transmission protocol utilized.
	In the following, we provide more concrete details for UTP and MMTP models.

	\subsection{Unicast Transmission Protocol (UTP)}
\begin{figure}[htbp]
		\includegraphics[width=8cm]{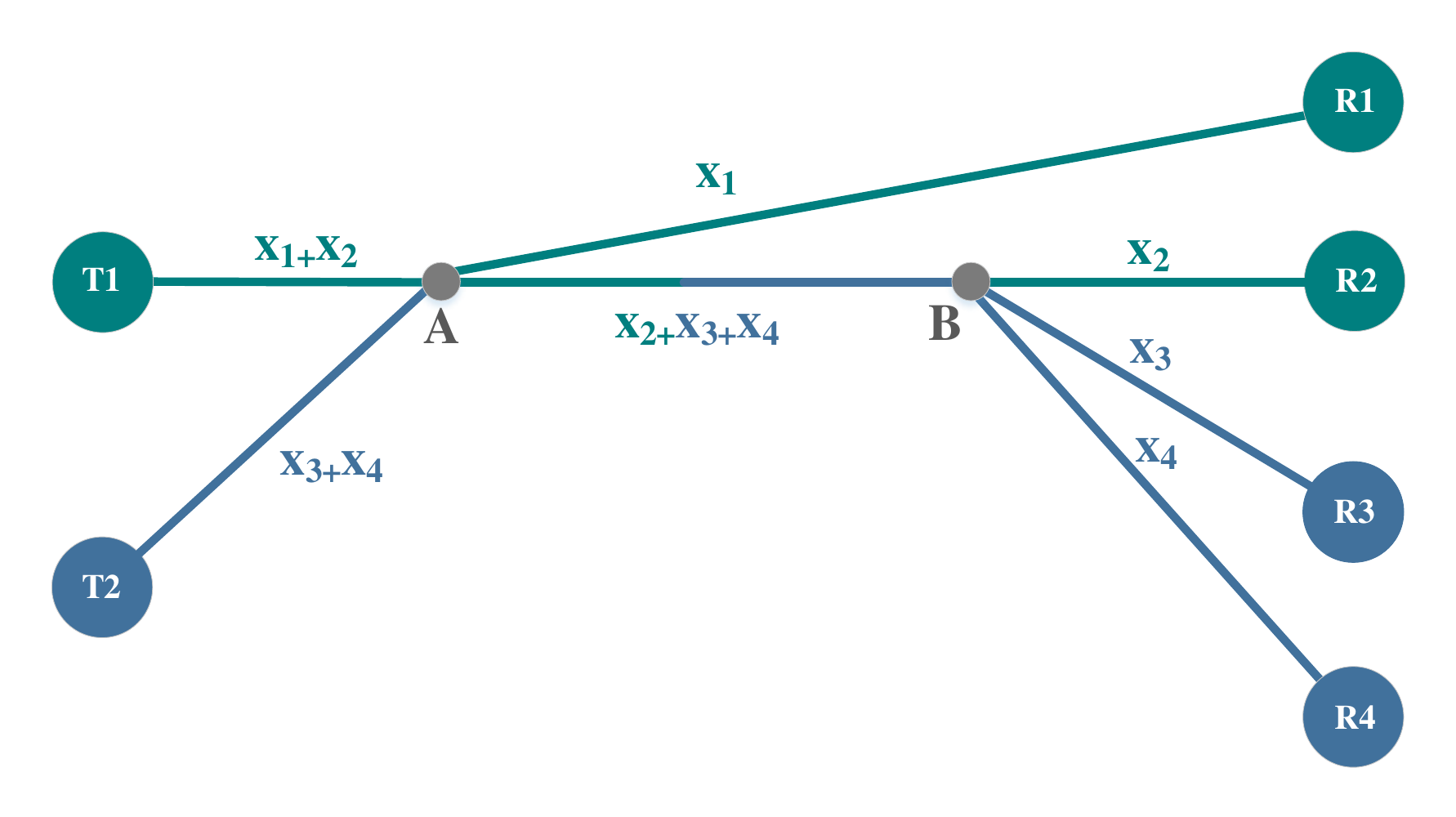}
		\caption{Network with Unicast Transmission Protocol (UTP).
			Both R1 and R2 use link T1-A, which, due to using UTP, is loaded with the sum data rate $x_1+x_2$. }
		\label{unicast}
	\end{figure}

	In UTP, a separate data stream is established between each source-receiver pair, regardless of whether the same data content is transmitted multiple times over some links. An example of a network utilizing UTP is depicted in Fig.~\ref{unicast}.
	We assume $N^l\geq 2$, that is at least two agents use any link $l$. This mild assumption is made so that there is competition between agents for using any of the links and it will help us avoid corner cases that distract from the main message of the paper.
	
	The centralized optimization problem that formulates the design goal for UTP is as follows
	\begin{subequations}\label{CPu}
		\begin{align}
		&\max_x \sum_{i \in \cN} v_i(x_i)  \label{CP1-a}\\
		\textrm{s.t.} \quad \ & x_i \geq 0 \quad \forall i \in \cN \label{CP1-b}\\
		\textrm{and} \quad \  & \sum_{j\in \cN^l} x_j \leq c^l \quad \forall l \in \cL  \label{CP1-c}.
		\end{align}
	\end{subequations}
	
	In order to characterize the solution of problem \eqref{CPu}, we use dual variables $\lambda=\{\lambda^l, l \in \cL\}$ and write the KKT conditions for this problem. Since the valuation functions are concave and all of the constraints of problem \eqref{CPu} are affine, problem \eqref{CPu} is a convex optimization problem and so KKT conditions are necessary and sufficient. These conditions at the optimal point $(x^*,\lambda^*)$ are
	
	\renewcommand{\labelenumi}{(\alph{enumi})}
	\begin{enumerate}
		\item Primal Feasibility: $x^*$ satisfies  \eqref{CP1-b} and  \eqref{CP1-c}.
		\item Dual Feasibility: $\lambda^{l*}\geq 0 \quad \forall l\in\cL.$
		\item Complimentary Slackness:
		\begin{subequations}\label{kktu}
			\begin{equation}
			\lambda^{l*}(c^l-\sum_{j\in\cN^l}x_j^*)=0 \quad \forall l\in\cL.
			\end{equation}
			\item Stationarity:
			\begin{align}
			v_i^\prime(x_i^*)&=\sum_{l\in\cL_i}\lambda^{l*} \quad \textrm{if} \quad x_i^*> 0\\
			v_i^\prime(x_i^*) &\leq \sum_{l\in\cL_i}\lambda^{l*} \quad \textrm{if}\quad x_i^*= 0.
			\end{align}
		\end{subequations}
	\end{enumerate}

	\subsection{Multirate Multicast Transmission Protocol (MMTP)}
	
\begin{figure}[h]
		\includegraphics[width=8cm]{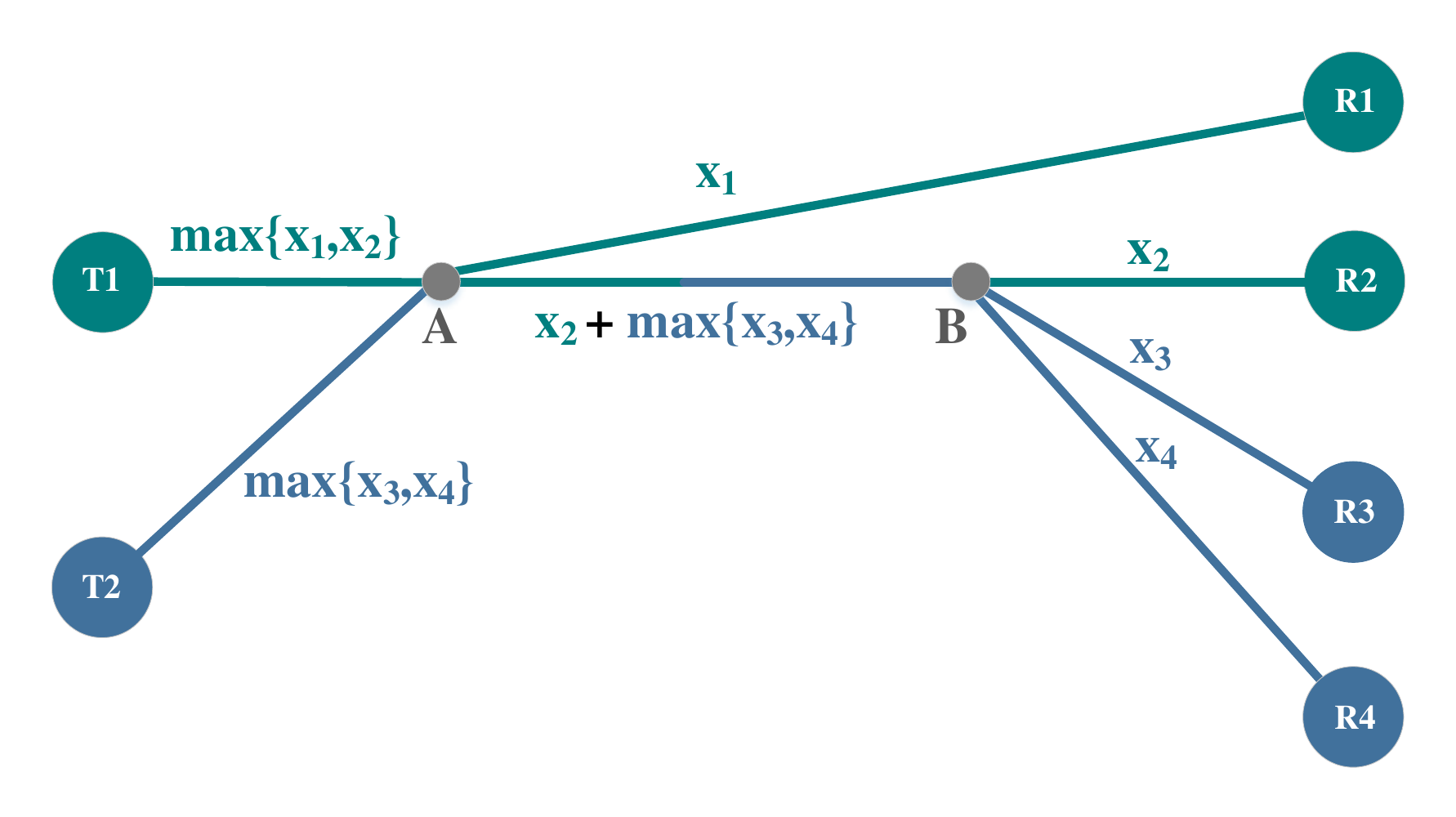}
		\caption{Network with Multirate Multicast Transmission Protocol (MMTP).
			Even though both R1 and R2 use link T1-A, due to using MMTP, it is only loaded with the maximum rate $\max\{x_1,x_2\}$.}
		\label{multicast}
	\end{figure}

	In MMTP model, agents are classified into $K$ groups based on their data source and each group is denoted by $\mathcal{G}_k\subset  \cN$.
	Since agents in each group receive data from the same source (albeit with possibly different data rates), instead of transmitting a separate data stream to each agent, in each link only a single stream is transmitted for each group utilizing that link. Furthermore, the data rate of that stream is the maximum demanded rate among users in the group on that link.
	In other words, each source transmits the common data of agents by the best quality demanded in each link and  each agent can regenerate her own data by down-sampling from the received data stream to get her desirable quality.
	This scenario is as if agents inside a group share the bandwidth with each other (public good) but they have competition for it with other groups (private good), a situation referred to as intergroup competition and intragroup sharing in~\cite{SiAn17b}.
	An example of a network utilizing MMTP is illustrated in Fig.~\ref{multicast}.
	We further define the following quantities.  The group of agent $i$ is denoted by $k (i)$. The set of users in group $k$ using link $l$ is denoted by  $\mathcal{G}_k^l=\cN^l \cap \mathcal{G}_k$. Further, $\mathcal{K}^l$ is the subset of groups that are using link $l$ (groups that contain at least one agent using link $l$) with $|\mathcal{K}^l|=K^l$.
	For the same reason as in UTP, we assume $K^l\geq 2$, that is, at least two groups use each link $l$ and this is for ensuring competition at each link.

	The centralized optimization problem that formulates the design goal for MMTP is as follows
	\begin{subequations}\label{CPm0}
		\begin{align}
		&\max_x \sum_{i \in \cN} v_i(x_i)  \label{CP-a}\\
		\textrm{s.t.} \quad \ & x_i \geq 0 \quad \forall i \in \cN \label{CP-b}\\
		\textrm{and} \quad \  & \sum_{k\in \mathcal{K}^l} \max_{i \in \mathcal{G}_k^l} \{x_i\} \leq c^l \quad \forall l \in \cL  \label{CP-c}.
		\end{align}
	\end{subequations}

	As in the case of UTP, we utilize KKT conditions to characterize the solution of problem~\eqref{CPm0}.
	We first need to transform it into a convex optimization problem. Towards this goal, we introduce the variable $b_k^l$ for each $l \in \cL$ and $k \in \mathcal{K}^l$ that represents the maximum demanded rate of agents in group $k$ that use link $l$, which we refer to as the ``group rate''. It is straightforward to show the equivalence of problem~\eqref{CPm0}  with the one below
	\begin{subequations}\label{CPm}
		\begin{align}
		&\max_x \sum_{i \in \cN} v_i(x_i)\label{CP2-a}\\
		\textrm{s.t.} \quad \ & x_i \geq 0 \quad \forall i \in \cN \label{CP2-b}\\
		\textrm{and} \quad \  & \sum_{k\in \mathcal{K}^l} b_k^l \leq c^l \quad \forall l \in \cL  \label{CP2-c}
		\\
		\textrm{and} \quad \  & x_i\leq b_k^l \quad \forall  l \in \cL , \ k \in \mathcal{K}^l, \ i \in \mathcal{G}_k^l. \label{CP2-d}
		\end{align}
	\end{subequations}
	Similar to problem~\eqref{CPu}, problem~\eqref{CPm} is a convex optimization problem and hence, KKT conditions are necessary and sufficient for its solution. We use dual variables $\lambda=\{\lambda^l, l\in \cL\}$, each of which corresponds to one of the constraints in~\eqref{CP2-c} and $\mu=\{\mu_i^l, \forall l \in \cL, i \in \cN^l\}$, each of which corresponds to one of the constraints in~\eqref{CP2-d}. We can write the KKT conditions at the optimal point $(x^*, b^*, \lambda^*, \mu^*)$ as
	\renewcommand{\labelenumi}{(\alph{enumi})}
	\begin{enumerate}
		\item Primal Feasibility: $x^*$ and $b^*$ satisfy  \eqref{CP2-b} and  \eqref{CP2-c} and \eqref{CP2-d}.
		\item Dual Feasibility: $\lambda^{l*} \geq 0 \ \forall l\in\cL$ and $\mu_i^{l*} \geq 0 \  \forall l \in \cL, i \in \cN^l$.
		\item Complimentary Slackness:
		\begin{subequations} \label{kktm}
			\begin{align}
			\lambda^{l*}(c^l-\sum_{k\in \mathcal{K}^l} b_k^{l*} )&=0 \quad \forall l\in\cL\\
			\mu_i^{l*}(x^*_i-b_k^{l*})&=0 \quad \forall l \in \cL, k \in \mathcal{K} ^l, i \in \mathcal{G}_k^l.
			\end{align}
			\item Stationarity:
			\begin{align}
			v_i^\prime(x_i^*)&=\sum_{l\in\cL_i}\mu_i^{l*} \quad \forall i \in \cN \quad \textrm{if} \quad x_i^*>0\\
			v_i^\prime(x_i^*)&\leq\sum_{l\in\cL_i}\mu_i^{l*} \quad \forall i \in \cN \quad \textrm{if} \quad x_i^*= 0\\
			\lambda^{l*}&=\sum_{i \in \mathcal{G}_k^l} \mu_i^{l*} \quad \forall  l \in \cL, k \in \mathcal{K}^l.
			\end{align}
		\end{subequations}
	\end{enumerate}
	
	Note that from the above KKT conditions it is obvious that MMTP gives rise to the ``free-riding'' problem that is commonly encountered in public-goods problems~\cite[Sec. 11.C]{MaWhGr95}: if an agent $i$ on link $l$ is not requesting the highest rate among the agents in the same group $k(i)$, then his contribution $\mu_i^{l*}$ is zero in the ``price'' $\lambda^{l*}$ for this link and thus she can free-ride on the other agent(s) in the group requesting the highest rate.

	The optimization problems  \eqref{CPu} and \eqref{CPm} cannot be solved in a centralized manner because the valuation functions of agents are private information of the agents. In the next section, we present two distributed mechanisms that aim to reach the solution of optimization problems \eqref{CPu} and \eqref{CPm} in a decentralized fashion in the presence of strategic agents.

	\section{Distributed Mechanism} \label{section3}
	
	A mechanism consists of a message space $\cM_i$ for each agent  $i \in \cN$ giving rise to a total message space $\cM=\cM_1 \times \ldots, \times \cM_N$, and allocation and tax functions that are denoted by $\hat{x}_i:\cM\rightarrow\mathbb{R}_+^N$ and $\hat{t}_i:\cM\rightarrow\mathbb{R}^N$, respectively.
	Hence, a mechanism can be characterized completely by specifying the tuple  $(\cM,(\hat{x}_i)_{i \in \cN},(\hat{t}_i)_{i \in \cN})$.
	The mechanism induces the game $\mathfrak{G}=( \cN, \cM, ( \hat{u}_i)_{i \in \cN} )$, where we consider a quasi-linear environment with $\hat{u}_i(m)=v_i(\hat{x}_i(m))-\hat{t}_i(m)$ for any $m\in\cM$.
	In the following we will use superscripts $\fu$ and $\fm$ to specify quantities for  UTP and MMTP, respectively. Thus, we will use notations $\cM^\fu$, $\hat{x}_i^\fu$, $\hat{t}_i^\fu$, $m^\fu$, $\mathfrak{G}^\fu$, $\hat{u}_i^\fu$ for UTP, and similarly, $\cM^\fm$, $\hat{x}_i^\fm$, $\hat{t}_i^\fm$, $m^\fm$, $\mathfrak{G}^\fm$, $\hat{u}_i^\fm$ for MMTP.
	In the following we formally describe ``distributed'' mechanisms, i.e., mechanisms for which the allocation and tax functions depend only on neighboring agents' messages as opposed to the entire message space $\cM$.

	\subsection{Message-Exchange Network}
	
	We describe the local exchange of messages through a ``message-exchange network", which is modeled as an undirected acyclic (tree) connected graph $\mathcal{GR=(N,E)}$, in which agents are denoted by nodes and an edge between two agents indicates that these two agents receive each others' messages. For all $i\in \cN$, $\cN(i)$ is the set of neighbors of agent $i$ in $\mathcal{GR}$ and $|\cN(i)|=N(i)$. Further, $n(i,j)$ is agent $i$'s neighbor which is on the shortest path from $i$ to $j$. Also, $\cN^l(i)=\cN(i)\cap \cN^l$ and $|\cN^l(i)|=N^l(i)$. For each agent $i \in \cN$, the function $\phi(i)$ arbitrarily chooses one of agent $i$'s neighbors. We define the set $\cI_i=\{h\in \cN(i):\phi(h)=i\}$. The role of this function will become evident in the description of the allocation and tax functions.
	
\begin{figure}[h]
		\includegraphics[width=8.5cm]{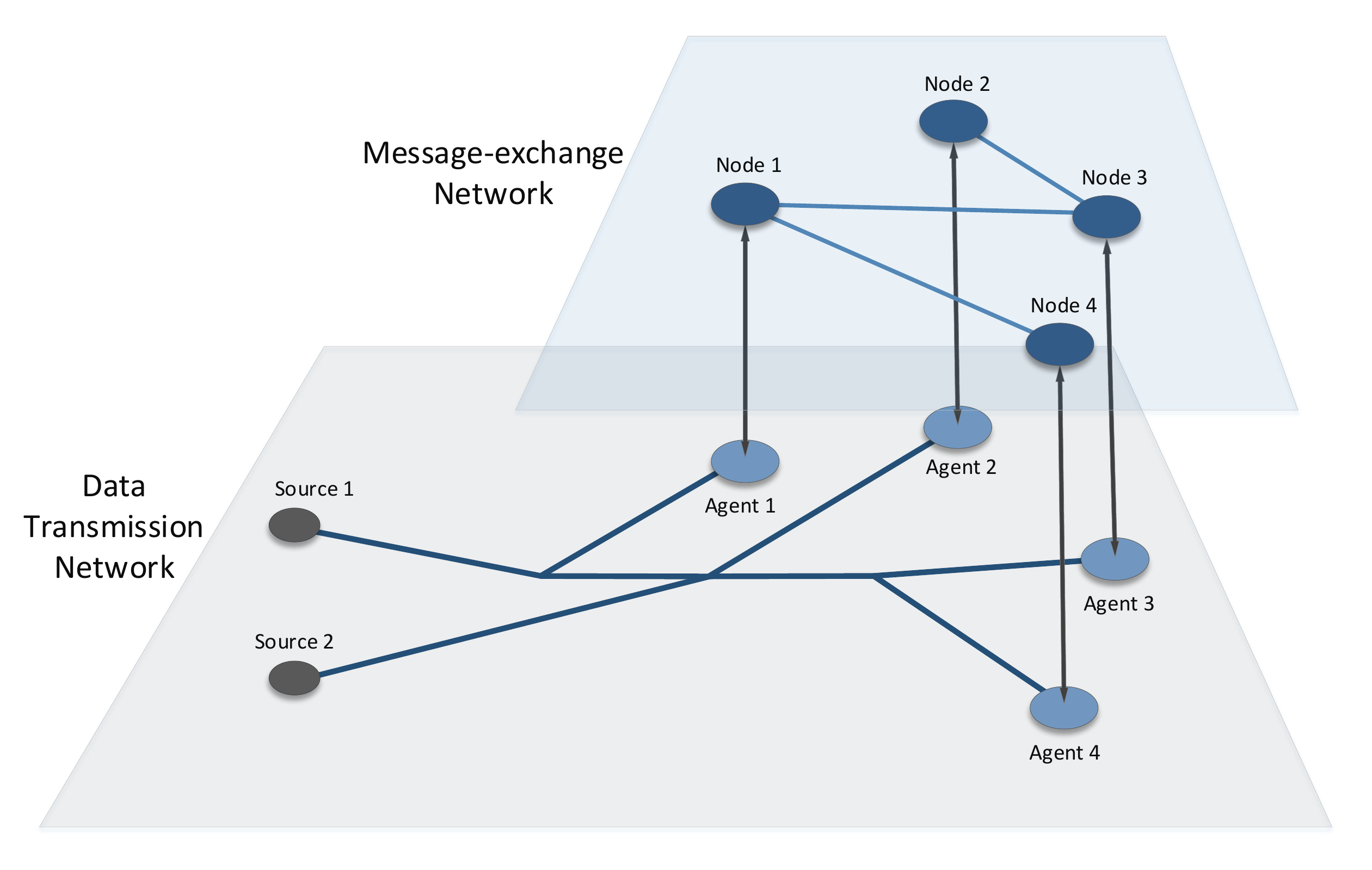}
		\caption{Message-exchange network vs. Data transmission network}
		\label{twolayer}
	\end{figure}

	Notice that the ``message-exchange network'' is not to be confused with the ``data transmission network'' related to UTP and MMTP and modeled through the centralized problems in \eqref{CPu} or \eqref{CPm0}. The former enables the decentralized solution of those problems by means of exchanging messages between neighboring nodes, while the latter describes the relation between agents dictated by the common links utilized by their data flows.
	These two neworks are illustrated in Fig.~\ref{twolayer}.

	In the following, we state two assumptions for the message-exchange network where Assumption~\ref{assump} holds for both UTP and MMTP mechanisms and Assumption \ref{assump2} only holds for MMTP mechanism. These assumptions simplify the exposition of the mechanisms. We will further relax Assumption~\ref{assump} for both mechanisms in Section~\ref{section5} and two alternative mechanisms will be proposed.
	
	\begin{assumption}\label{assump}
		(UTP/MMTP) For each link $l \in \cL$, the sub-graph consisting of agents $i \in   \cN^l$ is a connected graph.
	\end{assumption}
	
	This assumption essentially requires that a connected path exists for message passing between agents using the same link, eventually enabling them to form a consensus about some of the exchanged messages, e.g., the price for using each link.
	
	\begin{assumption}\label{assump2}
		(MMTP) For each link $l \in \cL$ and group $k \in \mathcal{K}^l$, there is at least one node $i \in \mathcal{G}_k^l$  that is connected to all other nodes $j \in \mathcal{G}_k^l$ in a single hop. This node will be referred to as the ``group leader'' of group $k$ in link $l$ and will be denoted by $c(k,l)$.
	\end{assumption}
	
	For each agent $i \in \cN$, the set $\mathcal{C}_i$ is defined as the set of links $l$ for which $c(k(i),l)=i$, i.e., this set contains all links for which agent $i$ is a group leader.
	
	The reason for this assumption is that in MMTP due to the free-riding problem, we require that there exists a user
	in each group $\mathcal{G}_k^l$ who has access to some necessary information about her group (e.g., group demand, group price) and can announce this information to the rest of the agents in $\mathcal{G}_k^l$.

	\subsection{Message Components}
	
	\subsubsection{UTP}
	Agent $i$ announces the message $m_i^\fu=(y_i,n_i,q_i,p_i)$.
	The first message $y_i \in \mathbb{R}_+$ is a proxy for her demanded rate.
	The second message, $n_i=(n_{i,j}^l, j\in\cN(i) , l\in \cL) \in \mathbb{R}_+^{L\times N(i)}$ consists of components $n_{i,j}^l$, which are referred to as ``summary" messages, each of which is a proxy for the sum of demands of the agents $h \in \cN^l$ with $n(i,h)=j$. In other words, $n_{i,j}^l$ is the sum of demanded rates for all users that are connected to $i$ through her neighbor $j$.
	These messages are required by agent $i$ in order to assess the total demand on each link she is using.
	The third message, $q_i=(q_{i,j}, j\in \cI_i) \in \mathbb{R}_+^{|\cI_i|}$ is a vector of components $q_{i,j}$, each of which is a proxy for the demand of neighboring agent $j \in \cI_i$. The purpose of these messages will become evident in the allocation function~\eqref{eq:allocations} and can intuitively be explained as follows: in order to allocate rate to agent $i$ in such a way that the capacity constraint at each link is satisfied, her demanded rate needs to be scaled by the sum of rates in that link. However, in evaluating the sum of rates, the rate of agent $i$ should not be controlled by her; instead an arbitrarily chosen neighboring agent $j$ quotes her rate through the message $q_{j,i}$. Clearly, we want the message $q_{j,i}$ to agree with $y_i$ at NE.
	Finally, the message $p_i=(p_i^l, l \in \cL_i) \in \mathbb{R}_+^{L_i}$ is the price (per unit of rate) that agent $i$ suggests for using each link $l$. This message is essentially a proxy for the dual variable $\lambda^{l*}$ that appears in the KKT conditions \eqref{kktu}.

	The message components for UTP are summarized in Table~\ref{tu}.
	\begin{table}[ht]
		\caption{Message components of agent $i \in \cN$ in UTP mechanism}
		\centering
		\begin{tabular}{c c  c }
			\hline\hline
			\begin{tabular}{c}Message \\ Component \end{tabular}  & Definition  & Functionality \rule{0pt}{3.6ex} \\ [1ex]
			\hline
			$y_i$ & -  & \begin{tabular}{c}  Demand for the data rate  \end{tabular} \rule{0pt}{3ex}  \\[1ex]  \hline
			$n_{i,j}^l$ &  \begin{tabular}{c} $j\in\cN(i)$ \\ $l\in \cL$ \end{tabular} & \begin{tabular}{c} Summary for demands of agents   \\   on  link $l$, connected to $i$ via $j$    \end{tabular} \rule{0pt}{3ex} \\  [1ex] \hline
			$q_{i,j}$ &   \begin{tabular}{c}$j\in \cI_i$ \end{tabular}  & \begin{tabular}{c} Proxy for  $y_j$  \end{tabular} \rule{0pt}{3ex}  \\ [1ex] \hline
			$p_i^l$ &  \begin{tabular}{c} $l\in \cL_i$ \end{tabular}  & \begin{tabular}{c} Suggested price  for using link $l$ \end{tabular} \rule{0pt}{3ex} \\
			[1ex]
			\hline
		\end{tabular}
		\label{tu}
	\end{table}

	\subsubsection{MMTP}
	
	Agent $i$ announces the message $m_i^\fm=(y_i,n_i,q_i,p_i,\yy_i,w_i,z_i,a_i)$.
	The reason for the larger message compared to UTP stems form the fact that (a) all agents within a group $\mathcal{G}_k^l$ need access to the maximum demanded rate in that group (this is required due to the free-riding problem that is inherent in the MMTP scenario) and (b) this information needs to be disseminated to all agents in the network while satisfying the communication constraints.
	In the following we give intuitive explanations for the meaning of each of the eight message components.
	The first message $y_i \in \mathbb{R}_+$ is a proxy for the agent $i$'s demanded rate.
	The second message, $n_i=(n_{i,j}^l, j\in\cN(i) , l\in \cL) \in \mathbb{R}_+^{L\times N(i)}$ consists of components $n_{i,j}^l$ where, similar to UTP, are referred to as ``summary'' messages, with the only difference being that each of them is a proxy for the sum of group demands (not individual demands) of the agents $h \in \cN^l$  with $n(i,h)=j$.
	The third message,  $q_i=(q_{i,j}, j \in \cI_i)\in \mathbb{R}_+^{|\cI_i|}$ consists of elements $q_{i,j}$, each of which is a proxy for agent $j$'s demand, and its role is similar to UTP.
	The fourth message consists of two components $p_i=(p_i^1,p_i^2)$. The first component is defined as $p_i^1=(p_i^{1,l}, l \in \cL_i) \in \mathbb{R}_+^{L_i}$, where each variable $p_i^{1,l}$ is the price that agent $i$ is willing to pay for using link $l$ and is similar to the message $p_i^l$ in UTP.
	This is essentially a proxy for the dual variable $\mu_i^{l*}$ that appears in the KKT conditions~\eqref{kktm}.
	The second component is defined as $p_i^2=(p_{i,j}^{2,l}, j\in \cI_i, l \in \cL_j) \in \mathbb{R}_+^{(\sum_{j\in \cI_i} L_j)}$, where each variable $p_{i,j}^{2,l}$ is the price that agent $i$ thinks agent $j$ should pay for using link $l \in \cL_j$.
	The reason why user $i$ quotes a price relevant to user $j$ is the same as in the case of the $q$ messages explained above in the UTP scenario.
	We now define the new messages that are present in MMTP and give intuitive explanations for their role.
	The fifth message is defined as $\yy_i=(\yy_i^l, l \in \cL_i)\in \mathbb{R}_+^{L_i}$, where, each of the variables $\yy_i^l$ is capturing whether the specific agent belongs in the group of agents that demand the maximum rate within the group $\mathcal{G}^l_{k(i)}$.
	We call these messages as proxies of the ``group demand''. Specifically, at NE, the message $\yy_i^l$ becomes zero if agent $i$ is not in the max group in link $l$, and otherwise, it will be the maximum demanded rate of group $\mathcal{G}^l_{k(i)}$ divided by the number of users in the max group in link $l$.
	The sixth message $w_i=(w_i^l, l\in \cL_i) \in \mathbb{R}_+^{L_i}$ consists of components $w_i^l$, each of which is a proxy for the price that group $k(i)$ is willing to pay for link $l$ and will be refered to as group price of group $k(i)$ for link $l$. These messages have to converge at NE to the dual variables ${\lambda_l}^*$ in the KKT conditions~\eqref{kktm} for all users $i\in \cN^l$.
	The seventh message is defined as $z_i=(z_i^1,z_i^2)$. The first component $z_i^1=(z_i^{1,l}, \ l \in \mathcal{C}_i)\in \mathbb{R}_+^{|\mathcal{C}_i|}$ consists of elements $z_i^{1,l}$, each of which is a proxy for the maximum value of demands of agents in $\mathcal{G}_{k(i)}^l$, i.e., the total group demand of agents in the group $\mathcal{G}_{k(i)}^l$.
	Further, $z_i^2=(z_i^{2,l}, \ l \in \mathcal{C}_i)\in \mathbb{R}_+^{|\mathcal{C}_i|}$ consists of elements $z_i^{2,l}$, each of which is a proxy for the number of agents that have maximum demand in $\mathcal{G}_{k(i)}^l$.
	These messages are quoted by user $i$ for every link for which $i$ is the group leader of $\mathcal{G}_{k(i)}^l$.
	Finally, the eighth message, $a_i=(a_i^1,{a_i^2})$, consists of two components. The first component is defined as $a_i^1=(a_i^{1,l}, l \in \mathcal{L}_i)\in \mathbb{R}_{++}^{L_i}$. Its role is quite technical and will become evident in the proof of efficiency of the NE of this mechanism. The second component, $a_i^2=(a_{i,j}^{2,l},  j\in \cI_i, l \in \mathcal{L}_j) \in \mathbb{R}_{++}^{(\sum_{j\in \cI_i}L_j)}$, consists of the elements $a_{i,j}^{2,l}$, each of which is a proxy for the message $a_j^{1,l}$. The reason for user $i$ quoting such message relevant to user $j$ is the same as in the case of $q$ messages explained earlier in the UTP scenario.

	The message components for MMTP are summarized in Table~\ref{tm}.
	\begin{table}[ht]
		\caption{Message components of agent $i \in \cN$ in MMTP mechanism}
		\centering 	
		\begin{tabular}{c c c }
			\hline\hline
			\begin{tabular}{c}Message \\ Component \end{tabular}  & Definition &  Functionality  \rule{0pt}{3.6ex}\\ [1ex]
			\hline
			$y_i$ &  -  & \begin{tabular}{c}  Demand for the data rate  \end{tabular}\rule{0pt}{3ex}  \\ [1ex]  \hline
			$n_{i,j}^l$ &  \begin{tabular}{c} $j\in\cN(i)$ \\ $l\in \cL$ \end{tabular}  & \begin{tabular}{c} Summary for group demands     \\ of agents  on  link $l$  and \\ connected to $i$ via $j$  \end{tabular} \rule{0pt}{3ex} \\ [1ex]  \hline
			$q_{i,j}$ &   $j\in \cI_i$  & \begin{tabular}{c} Proxy for $y_j$  \end{tabular} \rule{0pt}{3ex} \\ [1ex]  \hline
			
			$p_i^{1,l}$ &   $l\in \cL_i$  & \begin{tabular}{c} Suggested price  for using link $l$ \end{tabular} \rule{0pt}{3ex} \\ [1ex] \hline
			$p_{i,j}^{2,l}$ &   \begin{tabular}{c}$j \in \cI_i$ \\ $l\in \cL_j$ \end{tabular}  & \begin{tabular}{c} Proxy for $p_j^{1,l}$ \end{tabular} \rule{0pt}{3ex} \\ [1ex] \hline
			
			$s_i^l$ &  \begin{tabular}{c} $l\in \cL_i$ \end{tabular}  & \begin{tabular}{c} Proxy for group demand \\ on link $l$  and group $k(i)$ \end{tabular} \rule{0pt}{3ex} \\   [1ex] \hline
			$w_i^l$ &  \begin{tabular}{c} $l\in \cL_i$ \end{tabular}  & \begin{tabular}{c} Proxy for group price\\ of group $k(i)$ for link $l$ \end{tabular} \rule{0pt}{3ex} \\ [1ex] \hline
			$z_i^{1,l}$ &   $l\in \mathcal{C}_i$  & \begin{tabular}{c} Proxy for the total group \\ demand on link $l$  and group $k(i)$ \end{tabular} \rule{0pt}{3ex} \\[1ex] \hline
			$z_i^{2,l}$ &   $l\in \mathcal{C}_i$  & \begin{tabular}{c} Proxy for the number \\ of agents  with max demand \\ on link $l$ and group $k(i)$\end{tabular} \rule{0pt}{3ex} \\ [1ex] \hline
			$a_i^{1,l}$ &   $l\in \cL_i$  & \begin{tabular}{c} Technical point in the proof  \end{tabular} \rule{0pt}{3ex} \\ [1ex] \hline
			$a_{i,j}^{2,l}$ &   \begin{tabular}{c}$j \in \cI_i$ \\ $l\in \cL_j$ \end{tabular}  & \begin{tabular}{c} Proxy for $a_j^{1,l}$ \end{tabular} \rule{0pt}{3ex} \\ [1ex] \hline
		\end{tabular}
		\label{tm}
	\end{table}

	\subsection{Allocation Functions}
	
	Let us first define some auxiliary variables.
	For the UTP scenario, for each agent $i \in \cN$ and for every $l \in \cL$, $y_i^l$ is defined as $y_i^l=\textbf{1}_{\cL_i}(l) y_i$, where $\textbf{1}_{\mathcal{A}}$ is the indicator function of the set $\mathcal{A}$.
	Similarly, for the MMTP scenario, we define ${{y}}_i^l$ as $y_i^l=\textbf{1}_{\cL_i}(l) \yy_i^l$.
	Further, for each agent $i \in \cN$ and $l \in \cL_i$, the auxiliary quantities $ \bar{z}_i^{1,l}$ and $\bar{z}_i^{2,l}$ are defined as
	\begin{subequations}
		\begin{equation}
		\bar{z}_i^{1,l}=\left\{
		\begin{array}{cc}
		\max\{q_{\phi(i),i},\max_{j \in \mathcal{G}_{k(i)}^l,j \neq i}\{y_j\}\} & \quad \textrm{if} \ l \in \mathcal{C}_i \\
		z_{c(k(i),l)}^{1,l} & \quad \textrm{if} \ l \notin \mathcal{C}_i
		\end{array}\right.
		\end{equation}
		\begin{equation}
		\bar{z}_i^{2,l}=\left\{
		\begin{array}{cc}
		\textbf{1}_{\{q_{\phi(i),i}\}}( \bar{z}_i^{1,l})+\sum_{j \in \mathcal{G}_{k(i)}^l,j \neq i}\textbf{1}_{\{y_j\}}(\bar{z}_i^{1,l}) & \ \textrm{if} \ l \in \mathcal{C}_i \\
		z_{c(k(i),l)}^{2,l} & \ \textrm{if} \ l \notin \mathcal{C}_i.
		\end{array}\right.
		\end{equation}
	\end{subequations}
	
	The meaning of these quantities is as follows.
	The quantity $\bar{z}_i^{1,l}$ encodes the maximum demanded rate in the group $\mathcal{G}_{k(i)}^l$. If user $i$ is not the leader of that group, then the leader $c(k(i),l)$ quotes this message through $z_{c(k(i),l)}^{1,l}$.
	On the other hand, if $i$ is the leader of the group then she has to compute the maximum demand from all other members of the group including her own demand which is now quoted by a proxy through $q_{\phi(i),i}$.
	Similarly, the quantity $\bar{z}_i^{2,l}$ encodes the number of agents with maximum demand among all of the agents in the group $\mathcal{G}_{k(i)}^l$.

	We utilize an idea similar to the radial allocation \cite{SiAn17b} to have feasible allocation at NE. With this goal in mind, for message vectors $m_i^\fu=(y_i,n_i,q_i,p_i)$ and $m_i^\fm=(y_i,n_i,q_i,p_i,\yy_i,w_i,z_i,a_i)$, the allocation functions for the two mechanisms are defined as appropriately scaled versions of the demanded rates as follows
	\begin{subequations} \label{eq:allocations}
		\begin{align}\label{eq:allocations_radial}
		\hat{x}_i^\fu(m^\fu)=r_i^\fu \  {y_i}\ \ \ \ \
		\hat{x}_i^\fm(m^\fm)=r_i^\fm \  {y_i},
		\end{align}
		where $r_i^\fu$ and $r_i^\fm$ are agent $i$'s radial allocation factor in the two protocols and they are defined as
		\begin{align}
		r_i^\fu=\min_{l\in\cL} \frac{c^l}{f_i^{\fu,l}} \ \ \ \ \
		r_i^\fm=\min_{l\in\cL} \frac{c^l}{f_i^{\fm,l}},
		\end{align}
		where for $l \in \cL_i$, $f_i^{\fu,l}$ and $f_i^{\fm,l}$ are defined as
		\begin{align}
		f_i^{\fu,l}&=q_{\phi(i),i}+\sum_{j\in\cN(i)}(y_j^l+\sum_{h\in\cN(j) , h \neq i}n_{j,h}^l)\\
		f_i^{\fm,l}&=\frac{q_{\phi(i),i}\textbf{1}_{\{q_{\phi(i),i}\}}( \bar{z}_i^{1,l})}{\bar{z}_i^{2,l}}+\sum_{j\in\cN(i)}(y_j^l+\sum_{h\in\cN(j) , h \neq i}n_{j,h}^l),
		\label{eq:f_mmtp}
		\end{align}
		and for $l \notin \cL_i$,
		\begin{equation}
		f_i^{\fu,l}=f_i^{\fm,l}=\sum_{j\in\cN(i)}(y_j^l+\sum_{h\in\cN(j) , h \neq i}n_{j,h}^l).
		\end{equation}
	\end{subequations}
	The role of the messages $n$ and $q$ should now be clear from the above description. The quantity $f_i^{\fu,l}$ is the total demanded rate on link $l$ by all agents (from agent $i$'s viewpoint). However, since agent $i$ does not have access to other agents' demands outside her neighborhood, utilizes the summary messages $n_j$, $j\in\cN(i)$ for this purpose. In addition, her own demand is quoted by a pre-specified neighboring agent $\phi(i)$ through $q_{\phi(i),i}$. This is done so that the quantities $f^{\fu,l}_i$ and $f^{\fm,l}_i$ do not depend on agent $i$'s messages and the only part that agent $i$ can control in her allocation is $y_i$ in~\eqref{eq:allocations_radial}. This choice will greatly simplify our proofs of efficiency of the mechanisms.
	
	The additional complication in~\eqref{eq:f_mmtp} is due to the fact that in MMTP, if there are more than one agents who quote the maximum rate in a group, they should only load the corresponding link by that maximum rate and not the sum of the maximum rates, thus the normalization by $\bar{z}_i^{2,l}$. This is exactly the reason for the introduction of the $z$ messages in MMTP mechanism.

	\subsection{Tax Functions}
	\subsubsection{UTP}
	The tax function is $\hat{t}_i^\fu(m^\fu)=\sum_{l\in\cL}\hat{t}_i^{\fu,l}(m^\fu)$ and for each component $\hat{t}_i^{\fu,l}(m^\fu)$ we have different cases.
	
	For $l \in \cL_i$ we have
	\begin{subequations}
		\begin{align}\label{tax1}
		\hat{t}_i^{\fu,l}(m^\fu)=&\bar{p}_{-i}^l\hat{x}_i^\fu(m^\fu) \nonumber \\
		&+\sum_{j\in\cN(i)}(n_{i,j}^l-y_j^l-\sum_{h\in\cN(j) , h \neq i}n_{j,h}^l)^2 \nonumber \\
		&+\sum_{j\in \cI_i}(q_{i,j}-y_j)^2
		+(p_i^l-\bar{p}_{-i}^l)^2 \nonumber \\
		&+(p_i^l-\bar{p}_{-i}^l)\bar{p}_{-i}^l(c^l-r_i^\fu f_i^{\fu,l})^2,
		\end{align}
		where $\bar{p}_{-i}^l$ is the average price for link $l$ quoted by the neighbors of $i$ and  is defined as
		\begin{equation}
		\bar{p}_{-i}^l=
		\frac{1}{N^l(i)}\sum_{j \in \cN^l(i)}p_j^l.
		\label{p-i}
		\end{equation}
		For $l \notin \cL_i$ we have
		\begin{equation}
		\hat{t}_i^{\fu,l}(m^\fu)=\sum_{j\in\cN(i)}(n_{i,j}^l-y_j^l-\sum_{h\in\cN(j) , h \neq i}n_{j,h}^l)^2.
		\end{equation}
	\end{subequations}
	
	Intuitively, the tax functions provide some penalties to incentivize agents for quoting messages in a desirable way. With this goal in mind, taxes contain three types of terms.
	The first type is a rate times price component (first term in~\eqref{tax1}). Since agent $i$ controls her allocation through $y_i$ we do not allow her to control her price as well and so the price that she pays is dictated by her neighbors through $\bar{p}_{-i}^l$.
	The second type consists of quadratic terms that at NE will become zero and thus can be thought of as incentivizing agents to come to a consensus (second to fourth terms in~\eqref{tax1}). This enables the mechanism to provide proxies for the missing information of agents at NE, in addition to the requirements of having efficient allocation at NE.
	The third type relates to the complimentary slackness conditions in~\eqref{kktu} (fifth term in~\eqref{tax1}).
	The reason of defining two different tax functions is that different incentives are required for agents that utilize a link versus ones that do not. For instance, each agent $i$ has to pay a tax even for links $l \notin \cL_i$, which is required for consensus about the ``summary messages''. The intuition about each tax term will become more evident from the results in the lemmas of Section~\ref{section4}.

	\subsubsection{MMTP}
	The tax function is defined as $\hat{t}_i^\fm(m^\fm)=\hat{t}_i^{\fm,\mathfrak{c}}(m^\fm)+\sum_{l\in\cL}\hat{t}_i^{\fm,l}(m^\fm)$,
	where the first term is defined as
	\begin{subequations}
		\begin{align}
		\hat{t}_i^{\fm,\mathfrak{c}}(m^\fm)=&\sum_{j \in \cI_i}\sum_{l \in \cL_j}((p_{i,j}^{2,l}-{p_j^{1,l}})^2+(a_{i,j}^{2,l}-{a_j^{1,l}})^2) \nonumber \\
		&
		+\sum_{j\in \cI_i}(q_{i,j}-y_j)^2,
		\end{align}
		and for each component $\hat{t}_i^{\fm,l}(m^\fm)$, we consider different  cases.
		For $l \in \cL_i , l \notin \mathcal{C}_i$, we have
		
		\begin{align}\label{eq:t1}
		&\hat{t}_i^{\fm,l}(m^\fm)= {p_{\phi(i),i}^{2,l}}\hat{x}_i^\fm(m^\fm) \nonumber +\sum_{j\in\cN(i)}(n_{i,j}^l-y_j^l-\hspace{-0.3cm}\sum_{h\in\cN(j) , h \neq i}\hspace{-0.3cm}n_{j,h}^l)^2 \nonumber \\
		&+(\yy_i^l\hspace{-0.05cm}-\hspace{-0.05cm}\frac{q_{\phi(i),i}\textbf{1}_{\{q_{\phi(i),i}\}}( \bar{z}_i^{1,l})}{\bar{z}_i^{2,l}})^2
		\hspace{-0.1cm}+\bar{w}_{-i}^l(\hat{w}_i^l-\bar{w}_{-i}^l)(c^l\hspace{-0.1cm}-r_i^\fm f_i^{\fm,l})^2  \nonumber  \\
		&+(\hat{w}_i^l-\bar{w}_{-i}^l)^2+{p_{\phi(i),i}^{2,l}}(p_i^{1,l}- {p_{\phi(i),i}^{2,l}})( \bar{z}_i^{1,l}-q_{\phi(i),i})^2 \nonumber  \\
		&+(w_i^l-w_{c(k(i),l)}^l)^2.
		\end{align}
		
		For $l \in \cL_i , l \in \mathcal{C}_i$, we have
		\begin{align}
		&\hat{t}_i^{\fm,l}(m^\fm)= {p_{\phi(i),i}^{2,l}}\hat{x}_i^\fm(m^\fm) \nonumber+\sum_{j\in\cN(i)}(n_{i,j}^l-y_j^l-\hspace{-0.3cm}\sum_{h\in\cN(j) , h \neq i}\hspace{-0.3cm}n_{j,h}^l)^2 \nonumber \\
		&+(\yy_i^l-\frac{q_{\phi(i),i}\textbf{1}_{\{q_{\phi(i),i}\}}( \bar{z}_i^{1,l})}{\bar{z}_i^{2,l}})^2+(z_i^{1,l}- {\bar{z}_i^{1,l}})^2+(z_i^{2,l}-{\bar{z}_i^{2,l}})^2 \nonumber \\
		&+\bar{w}_{-i}^l(\hat{w}_i^l-\bar{w}_{-i}^l)(c^l-r_i^\fm f_i^{\fm,l})^2 +(\hat{w}_i^l-\bar{w}_{-i}^l)^2 \nonumber \\
		&+ {p_{\phi(i),i}^{2,l}}(p_i^{1,l}-{p_{\phi(i),i}^{2,l}})( \bar{z}_i^{1,l}-q_{\phi(i),i})^2 \nonumber \\
		&+ (w_i^l-{p_{\phi(i),i}^{2,l}}-\sum_{j \in \mathcal{G}_{k(i)}^l, j \neq i}{p_j^{1,l}})^2,
		\end{align}
		where for each link $l$ and agent $i \in \cN^l$, $\hat{w}_i^l$ is defined as
		\begin{equation}
		\hat{w}_i^l=\left\{
		\begin{array}{cc}
		\sum_{j \in \mathcal{G}_{k(i)}^l}{p_j^{1,l}}+ (a_i^{1,l}-{a_{\phi(i),i}^{2,l}}) & \hspace{-0.2cm} \textrm{if} \ l \in \mathcal{C}_i\\[0.2cm]
		w_{c(k(i),l)}^l-{p_{\phi(i),i}^{2,l}}+{p_i^{1,l}}+(a_i^{1,l}-{a_{\phi(i),i}^{2,l}}) & \hspace{-0.2cm} \textrm{if} \ l \notin \mathcal{C}_i,
		\end{array}
		\right.
		\end{equation}
		and, $\bar{w}_{-i}^l$ is defined as
		\begin{equation}
		\bar{w}_{-i}^l=
		\frac{1}{N^l(i)}\sum_{j \in \cN^l(i)}w_j^l.
		\label{w-i}
		\end{equation}
		Finally for $l \notin \cL_i$, the tax term is defined as
		\begin{equation} \hat{t}_i^{\fm,l}(m^\fm)=\sum_{j\in\cN(i)}(n_{i,j}^l-y_j^l-\sum_{h\in\cN(j) , h \neq i} n_{j,h}^l)^2.
		\end{equation}
	\end{subequations}

	The intuitive explanation of the terms appearing in the tax function is very similar to the one given above for the UTP scenario.
	The additional complication stems from the fact that we need to keep track of two types of prices, $p$ and $w$, corresponding to dual variables $\mu$ and $\lambda$, respectively.
	

	\section{A concrete example with UTP}\label{section_example}
	
	In this section, we provide a simple but not trivial example of UTP model and the corresponding mechanism for that. Assume that we have three agents $\cN=\{1,2,3\}$ using a single link (link $1$) with capacity $c^1=1$. The valuation function of agent $i \in \cN$ is given by $v_i(x_i)=i \ln(x_i)$. The optimization problem \eqref{CPu} will be of the form
	\begin{subequations}\label{CPuex}
		\begin{align}
		\max_x  \quad \ &  \ln(x_1)+2\ln(x_2)+3\ln(x_3) \\
		\textrm{s.t.} \quad \ & x_i \geq 0 \quad \forall i \in \cN \\
		\textrm{and} \quad \  & x_1+x_2+x_3 \leq 1.
		\end{align}
	\end{subequations}
	
	By writing the KKT conditions for this problem, one can easily calculate the solution to this optimization problem to be $x^*=(\frac{1}{6},\frac{1}{3},\frac{1}{2})$ and the optimal dual variable is $\lambda=6$. We will show that the mechanism for UTP, has Nash equilibria, all of which result in this efficient allocation $x^*$.
	
\begin{figure}[ht]
		\centering
		\includegraphics[width=5cm]{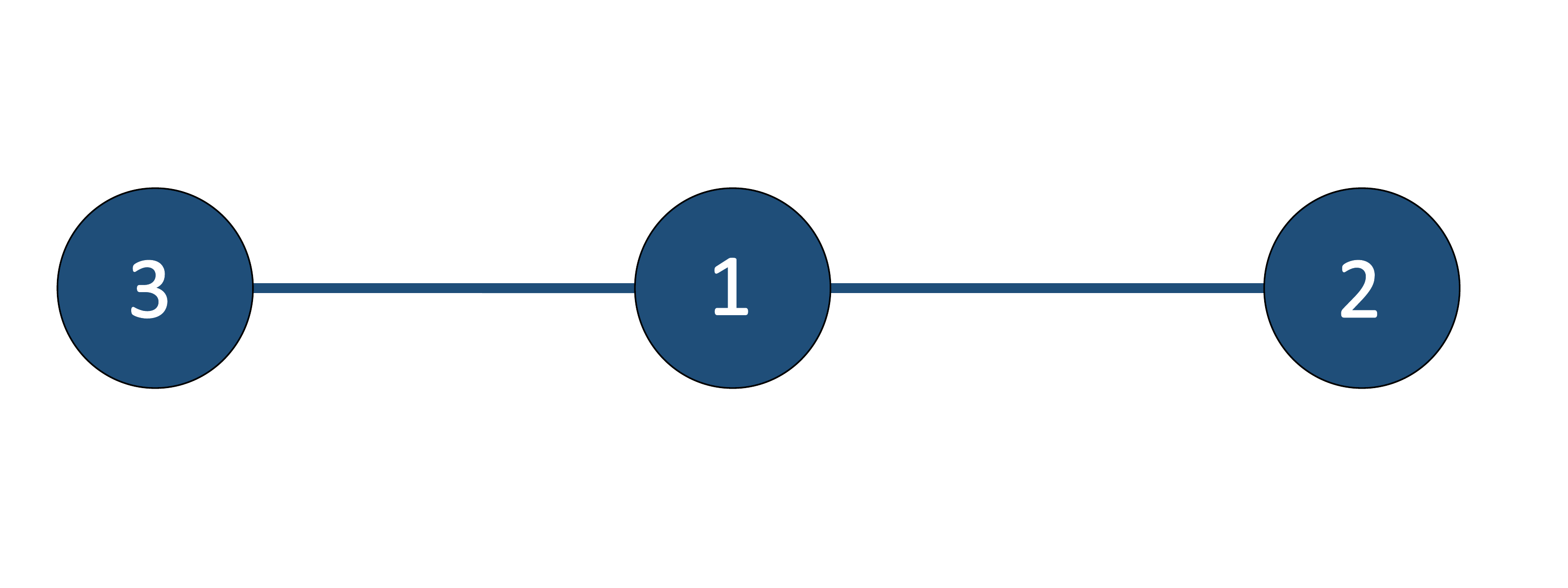}
		\captionsetup{justification=centering}
		\caption{Message-exchange network }
		\label{mesnetex}
	\end{figure}

	We consider the message network of Figure \ref{mesnetex} between the agents. Note that this message network satisfies Assumption \ref{assump} and it is a tree graph.
	We know that $\phi(2)=1$ and $\phi(3)=1$ (they only have one option). Assume that $\phi(1)=2$.
	The above means that agent~1 uses agent~2 for quoting a proxy of his demand,
	while agent~2 uses agent~1 and agent~3 uses agent~1 for the same.
	The message components of agents are $m_1^\fu=(y_1,p_1^1,n_{1,2}^1,n_{1,3}^1,q_{1,2},q_{1,3})$, $m_2^\fu=(y_2,p_2^1,n_{2,1}^1,q_{2,1})$ and $m_3^\fu=(y_3,p_3^1,n_{3,1}^1)$.
	In this simple, single-link setting, superscripts $^1$ are redundant but we maintain them for notational uniformity with the general description. In this network, agent~1 can listen to all messages, while agent~2 cannot listen to $m_3$, and similarly, agent~3 cannot listen to $m_2$.
	The allocation functions are as follows
	\[
	\hat{x}_1^\fu(m^\fu)=r_1^\fu y_1 \quad \hat{x}_2^\fu(m^\fu)=r_2^\fu y_2 \quad \hat{x}_3^\fu(m^\fu)=r_3^\fu y_3,
	\]
	where
	\begin{align*}
	&r_1^\fu=\frac{1}{q_{2,1}+y_2+y_3}\\
	&r_2^\fu=\frac{1}{y_1+q_{1,2}+n_{1,3}^1} \\
	&r_3^\fu=\frac{1}{y_1+n_{1,2}^1+q_{1,3}}.
	\end{align*}
	Observe the roles of the $q$ and the $n$ message components. All agents would have to scale their messages by the total demand $y_1+y_2+y_3$. We do not want agent~1 to control her scaling factor and thus we ask agent~2 to quote a proxy $q_{2,1}$ for her demand $y_1$. A similar argument for agents~2 and 3 justifies the presence of the messages $q_{1,2}$ and $q_{1,3}$. In addition, agent~2 does not have access to the demand quoted by agent~3 and that's why she is using the summary message $n^1_{1,3}$ quoted by agent~1 for this purpose. Similarly for agent~3.
	Finally, note that summary messages $n^1_{2,1}$ and $n^1_{3,1}$ are redundant and are not used in this example. Since agents~2 and~3 are at the leafs of the tree they do not need to pass any information downstream, so these messages are not used in the mechanism.
	
	The tax functions can be written as follows
	\begin{align*}
	&t_1(m^\fu)=\bar{p}_{-1}^1 \hat{x}_1^\fu\\
	&+(n_{1,2}^1-y_2)^2+(n_{1,3}^1-y_3)^2+(q_{1,2}-y_2)^2+(q_{1,3}-y_3)^2\\
	&+(p_1^1-\bar{p}_{-1}^1)^2+(p_1^1-\bar{p}_{-1}^1)\bar{p}_{-1}^1(1-r_1^\fu (q_{2,1}+y_2+y_3) )^2\\
	&t_2(m^\fu)=\bar{p}_{-2}^1 \hat{x}_2^\fu+(n_{2,1}^1-y_1-n_{1,3}^1)^2+(q_{2,1}-y_1)^2\\
	&+(p_2^1-\bar{p}_{-2}^1)^2+(p_2^1-\bar{p}_{-2}^1)\bar{p}_{-2}^1(1-r_2^\fu (y_1+q_{1,2}+n_{1,3}^1) )^2\\
	&t_3(m^\fu)=\bar{p}_{-3}^1 \hat{x}_3^\fu+(n_{3,1}^1-y_1-n_{1,2}^1)^2\\
	&+(p_3^1-\bar{p}_{-3}^1)^2+(p_3^1-\bar{p}_{-3}^1)\bar{p}_{-3}^1(1-r_3^\fu (y_1+n_{1,2}^1+q_{1,3}))^2,
	\end{align*}
	where
	$$\bar{p}_{-1}^1=\frac{p_2^1+p_3^1}{2} \quad \bar{p}_{-2}^1=p_1^1 \quad \bar{p}_{-3}^1=p_1^1.$$
	Since the $n$ and $q$ messages only appear once in the tax function of each agent, each agent has the ability to minimize the tax terms by zeroing out the corresponding quadratic terms (four terms for agent~1, two terms for agent~2 and one term for agent~3). So, at NE, we have
	\begin{align*}
	&n_{1,2}^1=y_2 \quad n_{1,3}^1=y_3 \quad n_{2,1}^1=y_1+n_{1,3}^1 \quad n_{3,1}^1=y_1+n_{1,2}^1 \\ &q_{1,2}=y_2 \quad q_{1,3}=y_3 \quad q_{2,1}=y_1.
	\end{align*}
	This means that at NE, we have
	\begin{align*}
	r_1^\fu=r_2^\fu=r_3^\fu=\frac{1}{y_1+y_2+y_3},
	\end{align*}
	and therefore,
	\begin{align*}
	\hat{x}_i^\fu(m^\fu)=\frac{y_i}{y_1+y_2+y_3}, \qquad i\in\cN.
	\end{align*}
	The above further implies that at NE, $\hat{x}_1^\fu(m^\fu)+\hat{x}_2^\fu(m^\fu)+\hat{x}_3^\fu(m^\fu)=1$ and the link is fully loaded, and as a result, the last term in all three tax functions will be zero.
	Consequently, each agent has now the ability to minimize the tax terms by zeroing out the remaining quadratic term that depends on the quoted price.
	We can conclude that at NE, $p_1^1=\bar{p}_{-1}^1, \  p_2^1=\bar{p}_{-2}^1=p_1^1, \ p_3^1=\bar{p}_{-3}^1=p_1^1$, which implies that all price messages are equal (to some yet unspecified price $p$) at NE, i.e., $p_1^1=p_2^1=p_3^1=p$.
	
	Furthermore, by deriving the best response for the messages $y_1, y_2$ and $y_3$ we have the following equation for $i \in \cN,$
	\begin{align*}
	&\frac{du_i(m^\fu)}{dy_i}=0 \quad \text{if}\  y_i>0\\
	&\frac{du_i(m^\fu)}{dy_i}<0 \quad \text{if} \ y_i=0,
	\end{align*}
	which implies that
	\begin{align*}
	&v_i'(\hat{x}_i^\fu)=p \quad \text{if} \  y_i>0\\
	&v_i'(\hat{x}_i^\fu)<p \quad \text{if} \ y_i=0.
	\end{align*}
	By solving these equations for $\hat{x}_i^\fu$ and $p$, we have $p=6$ (it equals to $\lambda$ in the optimization problem \eqref{CPuex}) and  $\hat{x}_i^\fu=\frac{i}{6}$ which means that $\hat{x}_1^\fu=\frac{1}{6}$, $\hat{x}_2^\fu=\frac{2}{6}$ and $\hat{x}_3^\fu=\frac{3}{6}$. Hence, the allocation at NE is efficient. The equilibrium $y$ messages can be derived from the following equation
	\begin{align*}
	\frac{y_i}{y_1+y_2+y_3}=\frac{i}{6},
	\end{align*}
	which implies that $y_i=k\frac{i}{6}$ for any constant number $k>0$. This shows that there are infinitely many NE, but all of them have the same and efficient allocation.

	\section{Mechanism Properties}  \label{section4}
	
	\begin{fact}\label{dist}
		The mechanisms  $\footnotesize{(\cM^\fu,\hat{x}^\fu,\hat{t}^\fu)}$ and $\footnotesize{(\cM^\fm,\hat{x}^\fm,\hat{t}^\fm)}$ are distributed.
	\end{fact}
	
	This can be obviously derived from the definition of allocation and tax functions. Clearly, they depend only on each agent's own messages and the messages of her neighboring agents.
	
	\begin{theorem}(Full Implementation, Individual Rationality and Weak Budget Balance - UTP)\label{FIu}
		The game $\mathfrak{G}^\fu$ has infinitely many Nash equilibria. At every Nash equilibrium $m^\fu \in \mathcal{NE}^\fu$ of the game $\mathfrak{G}^\fu$, the allocation vector $\hat{x}^\fu(m^\fu)$ is efficient, i.e., it is equal to the solution, $x^*$, of problem~\eqref{CPu}. In addition, for each agent, individual rationality is satisfied at all NE. Further, the game $\mathfrak{G}^\fu$ is weak budget balanced at all NE, i.e., $\sum_{i \in \cN}\hat{t}^\fu_i(m^\fu)\geq 0$.
	\end{theorem}

	\begin{theorem}(Full Implementation, Individual Rationality and Weak Budget Balance - MMTP)\label{FIm}
		The game $\mathfrak{G}^\fm$ has infinitely many Nash equilibria. At every Nash equilibrium $m^\fm \in \mathcal{NE}^\fm$ of the game $\mathfrak{G}^\fm$, the allocation vector $\hat{x}^\fm(m^\fm)$ is efficient, i.e., it is equal to the solution, $x^*$, of problem~\eqref{CPm0}. In addition, for each agent, individual rationality is satisfied at all NE. Further, the game $\mathfrak{G}^\fm$ is weak budget balanced at all NE, i.e., $\sum_{i \in \cN}\hat{t}^\fm_i(m^\fm)\geq 0$.
	\end{theorem}
	
	Regarding the multiplicity of Nash equilibria in the induced games, we point out that there are two reasons for this behavior.
	The first reason of not having a unique Nash equilibrium is that the dual variables are not generally unique and therefore,
	for each dual variable solution (price messages in the mechanism), we can construct a different Nash equilibrium.
	The second reason is that the demand vector in each of the Nash equilibria is a scaled version of the efficient allocation and every Nash equilibrium corresponds to a scaled version of the allocation.
	However, the resulting  allocation in all of these Nash equilibria is efficient as stated in the theorems.
	Since for both problems \eqref{CPu} and \eqref{CPm0} the solution is unique, then, according to Theorems \ref{FIu} and \ref{FIm}, for all $m^\fu \in \mathcal{NE}^\fu$, the allocation vector $\hat{x}^\fu(m^\fu)$ is unique, and for all $m^\fm \in \mathcal{NE}^\fm$, the allocation vector  $\hat{x}^\fm(m^\fm)$ is unique.

	Before proving Theorems \ref{FIu} and \ref{FIm}, some lemmas are presented that are necessary for their proof.
	The basic idea behind the proof is to show, through a series of lemmas, certain necessary conditions that all NE of the induced games $\mathfrak{G}^\fu$ and $\mathfrak{G}^\fm$ should satisfy. These necessary conditions essentially lead to showing that
	the allocations and prices at NE are satisfying the KKT conditions for problems~\eqref{CPu} and~\eqref{CPm}, respectively.
	The proof is concluded by showing that indeed there exists such an equilibrium by constructing it.

	\begin{lemma}(Concavity)\label{con}
		The function $\hat{u}^\fu_i(m_i^\fu,m_{-i}^\fu)$ is strictly concave w.r.t. $m_i^\fu$.
		Similarly, the function $\hat{u}^\fm_i(m_i^\fm,m_{-i}^\fm)$ is strictly concave w.r.t. $m_i^\fm$.
	\end{lemma}
	The strict concavity of $\hat{u}^\fu_i(m_i^\fu,m_{-i}^\fu)$ and $\hat{u}^\fm_i(m_i^\fm,m_{-i}^\fm)$ w.r.t. $m_i^\fu$ and  $m_i^\fm$, respectively, helps us calculate the best response functions for player $i$ in each of the games $\mathfrak{G}^\fu$ and $\mathfrak{G}^\fm$ by setting the gradient of  $\hat{u}^\fu_i(m_i^\fu,m_{-i}^\fu)$  w.r.t. $m_i^\fu$ and $\hat{u}^\fm_i(m_i^\fm,m_{-i}^\fm)$  w.r.t. $m_i^\fm$  to be equal to zero, respectively. Whenever an element of the gradient cannot be set to zero, it is either always positive or always negative. If any of the elements is always positive, then as message spaces are unbounded from above, there is no best response. Otherwise, if any of the elements of the gradient vector is always negative, the best response would be zero for that element.

	The next two lemmas are related with the quadratic terms in the tax functions of UTP and MMTP mechanisms.
	As mentioned earlier, at NE, agents force these quadratic terms to zero thus achieving consensus.
	For instance, each message component $q_{i,j}$ can be used as a proxy for $y_j$ by agent $j$ and yet is not controlled by agent $j$.
	Furthermore, these lemmas explain how summary messages $n$ are designed to sum up the demands (UTP) or group demands (MMTP) of all agents using link $l$ at NE.
	
	\begin{lemma}\label{qij1}
		At any $m\in \mathcal{NE}^\fu$ we have
		\begin{subequations}
			\begin{equation}
			q_{i,j}=y_j,\quad \forall j \in \cI_i,
			\end{equation}
			and
			\begin{equation}
			n_{i,j}^l=y_j^l+\sum_{h\in\cN(j) , h \neq i}n_{j,h}^l.
			\end{equation}
			This implies that at any NE,
			\begin{equation}\label{eq:summaries}
			n_{i,j}^l=\sum_{h\in\cN , n(i,h)=j}y_h^l.
			\end{equation}
		\end{subequations}		
	\end{lemma}
	Regarding~\eqref{eq:summaries}, note that since the message graph is a tree, each node is connected to any other node only by one path, and therefore, the demand of each agent is counted once. This ensures no double counting of demands.
	

	\begin{lemma}\label{quadratic}
		At any $m^\fm \in \mathcal{NE}^\fm$, the following equations hold for any $i \in \cN$,
		\begin{subequations}
			\begin{align}
			q_{i,j}=y_j, \ \forall j \in \cI_i
			\end{align}
			\begin{align}
			\yy_i^l=\frac{q_{\phi(i),i}\textbf{1}_{\{q_{\phi(i),i}\}}( \bar{z}_i^{1,l})}{\bar{z}_i^{2,l}}, \ \forall l \in \cL_i
			\end{align}
			\begin{align}		
			p_{i,j}^{2,l}={p_j^{1,l}} , \ \forall  j \in \cI_i, l \in \cL_j
			\end{align}
			\begin{align}	
			w_i^l=\left\{
			\begin{array}{cc}
			w_{c(k(i),l)}^l & \ \textrm{if} \quad  l \in \cL_i , l \notin \mathcal{C}_i\\
			{p_{\phi(i),i}^{2,l}}+\sum_{j \in \mathcal{G}_{k(i)}^l, j \neq i}{p_j^{1,l}} & \ \textrm{if} \quad  l \in \cL_i , l \in \mathcal{C}_i
			\end{array}\right.
			\end{align}
			\begin{align}
			n_{i,j}^l=y_j^l+\sum_{h\in\cN(j) , h \neq i}n_{j,h}^l , \ \forall l \in \cL , j \in \cN(i)
			\end{align}
			\begin{align}
			z_i^{1,l}= {\bar{z}_i^{1,l}}=\max\{q_{\phi(i),i},\max_{j \in \mathcal{G}_{k(i)}^l,j \neq i}y_j\} , \ \forall l \in \cL_i , l \in \mathcal{C}_i
			\end{align}
			\begin{align}
			&z_i^{2,l}= {\bar{z}_i^{2,l}}=\textbf{1}_{\{q_{\phi(i),i}\}}(\bar{z}_i^{1,l}) \nonumber \\
			&\qquad\qquad +\sum_{j \in \mathcal{G}_{k(i)}^l,j \neq i} \textbf{1}_{\{y_j\}}(\bar{z}_i^{1,l}), \ \forall l \in \cL_i, l\in \mathcal{C}_i
			\end{align}
			\begin{align}
			a_{i,j}^{2,l}= {a_j^{1,l}} , \ \forall j \in \cI_i, l \in \mathcal{L}_j.
			\end{align}
		\end{subequations}
	\end{lemma}
	

	In the next lemma it is shown how radial allocation (whereby the actual allocation is a scaled version of the requested allocation by all agents) ensures feasibility at NE.
	\begin{lemma}(Primal Feasibility)\label{pfeas}
		At any $m^\fu \in \mathcal{NE}^\fu$, of the game $\mathfrak{G}^\fu$, the allocation vector $\hat{x}^\fu(m^\fu)$ is feasible for  problem \eqref{CPu}. Similarly, at any  $m^\fm \in \mathcal{NE}^\fm$, of the game $\mathfrak{G}^\fm$, the allocation vector $\hat{x}^\fm(m^\fm)$ is feasible for  problem \eqref{CPm0}.
	\end{lemma}
	

	The next two lemmas show how different agents form a consensus on the price variables for each link $l$.
	For example it is shown that all quoted prices $p_i^l$ end up being equal to a price $p^l$ at NE for the UTP scenario.
	This price will play the role of the dual variable $\lambda^l$ in the  KKT conditions~\eqref{kktu}.
	Similarly, for the MMTP scenario, it is shown how different groups form a consensus on the group prices $\hat{w}_i^l$ for each link $l$ which becomes equal to $w^l$ at NE.
	Furthermore, in both lemmas, equilibrium expressions are derived that resemble the complimentary slackness terms of the  KKT conditions in~\eqref{kktu} and~\eqref{kktm}.
	\begin{lemma}\label{pil1}
		At any $m^\fu \in \mathcal{NE}^\fu$, of the game  $\mathfrak{G}^\fu$,
		\begin{subequations}
			\begin{equation}\label{eq:eq_prices}
			p_i^l=p^l, \forall i \in \cN , l \in \cL_i.
			\end{equation} Also,
			\begin{equation}
			p^l(c^l-\sum_{i\in\cN^l}\hat{x}^\fu_i)=0 \quad \forall l\in \cL.
			\end{equation}
		\end{subequations}
	\end{lemma}


	\begin{lemma}\label{pw}
		At any  $m^\fm \in \mathcal{NE}^\fm$, of the game $\mathfrak{G}^\fm$, the following constraints hold for all $i \in \cN$ and $l \in \cL_i$,
		\begin{subequations}
			\begin{align}
			\hat{w}_i^l &=w^l \label{eqw}\\
			w^l(c^l-r_i^\fm \sum_{i \in \cN}y_i^l) &=0 \label{comw}\\
			p_i^{1,l}(y_i-{\bar{z}_i^{1,l}}) &=0. \label{comp}
			\end{align}
		\end{subequations}
	\end{lemma}

	The next two lemmas conclude the necessary conditions by showing that NE implies
	the stationary conditions in \eqref{kktu} and \eqref{kktm}.
	\begin{lemma}\label{stationarity1}(Stationarity - UTP)
		At any  $m^\fu \in \mathcal{NE}^\fu$, of the game $\mathfrak{G}^\fu$, the following constraints are satisfied,
		\begin{subequations}
			\begin{align}
			v_i^\prime(\hat{x}^\fu_i(m^\fu)) &= \sum_{l\in\cL_i}p^l \quad \textrm{if}\quad \hat{x}^\fu_i(m^\fu)> 0\\
			v_i^\prime(\hat{x}^\fu_i(m^\fu)) &\leq \sum_{l\in\cL_i}p^l \quad \textrm{if}\quad \hat{x}^\fu_i(m^\fu)= 0.
			\end{align}	
		\end{subequations}
		
	\end{lemma}


	\begin{lemma}\label{stationarity2}(Stationarity - MMTP)
		The following constraints hold at any $m^\fm \in \mathcal{NE}^\fm$, of the game $\mathfrak{G}^\fm$,
		\begin{subequations}
			\begin{align}
			v_i^\prime(\hat{x}^\fm_i(m^\fm)) &= \sum_{l\in\cL_i}{p_i^{1,l}} \quad \textrm{if}\quad \hat{x}^\fm_i(m^\fm)> 0\\
			v_i^\prime(\hat{x}^\fm_i(m^\fm)) &\leq \sum_{l\in\cL_i}{p_i^{1,l}} \quad \textrm{if}\quad \hat{x}^\fm_i(m^\fm)= 0.
			\end{align}	
		\end{subequations}
		\label{stationarity}	
	\end{lemma}
	

	As mentioned earlier, the overall strategy for proving our main result is to show that any NE satisfies the KKT conditions of the original problem and then showing that such an equilibrium indeed exists. This last step is shown in the following lemma.
	\begin{lemma}(Existence of NE)\label{exist}
		There exist infinitely many Nash equilibria  $m^\fu \in \mathcal{NE}^\fu$, for the game $\mathfrak{G}^\fu$.
		Also, there exist infinitely many Nash equilibria $m^\fm \in \mathcal{NE}^\fm$, for the game $\mathfrak{G}^\fm$.
	\end{lemma}


	The above series of lemmas is sufficient to prove the full implementation result for the two mechanisms for UTP and MMTP.
	Individual rationality and weak budget balance are shown separately in the following lemma.
	
	\begin{lemma}(Individual Rationality and Weak Budget Balance)\label{IR}
		At any NE of the games $\mathfrak{G}^\fu$ and $\mathfrak{G}^\fm$, individual rationality is satisfied
		\begin{subequations}
			\begin{align}
			v_i(\hat{x}^\fu_i(m^\fu))-\hat{t}^\fu_i(m^\fu)&\geq v_i(0), \ \forall i \in \cN\\
			v_i(\hat{x}^\fm_i(m^\fm))-\hat{t}^\fm_i(m^\fm)&\geq v_i(0), \ \forall i \in \cN.
			\end{align}
		\end{subequations}
		Furthermore, both of the mechanisms are weak budget balanced
		\begin{subequations}
			\begin{align}
			\sum_{i \in \cN}\hat{t}^\fu_i(m^\fu)&\geq 0\\
			\sum_{i \in \cN}\hat{t}_i^\fm(m^\fm)&\geq 0.
			\end{align}
		\end{subequations}
	\end{lemma}


	We are now ready to state the proofs of Theorems~\ref{FIu} and \ref{FIm}.
	\begin{IEEEproof}[Proof of Theorem \ref{FIu}]
		In the proof of Lemma \ref{exist}, we showed that the message associated  to the solution of problem \eqref{CPu}, $(x^*,\lambda^*)$, is a NE of the game $\mathfrak{G}^\fu$. Now, we want to prove that all of the NE of the game $\mathfrak{G}^\fu$ generate allocation and prices that are efficient for problem \eqref{CPu}.
		Consider any $m^\fu\in \mathcal{NE}^\fu$, if $\hat{x}^\fu(m^\fu)$ is used as the primal variables vector and $p=\{p^1, ..., p^L\}$ is used as the dual variables vector, all of the KKT conditions \eqref{kktu} are satisfied due to Lemmas~\ref{qij1}, \ref{pfeas}, \ref{pil1}, and \ref{stationarity1}. Therefore, $\hat{x}^\fu(m^\fu)=x^*$ for any  $m^\fu\in \mathcal{NE}^\fu$ and so, full implementation is obtained.
		Furthermore, Lemma~\ref{IR} proves individual rationality and weak budget balance at any NE of the game $\mathfrak{G}^\fu$.
	\end{IEEEproof}
	
	The proof of Theorem~\ref{FIm} is very similar to the proof of Theorem~\ref{FIu} and it is stated below for completeness.
	\begin{IEEEproof}[Proof of Theorem \ref{FIm}]
		Let $(x^*,b^*,\lambda^*,\mu^*)$ be the solution of problem~\eqref{CPm} and consider any $m^\fm\in \mathcal{NE}^\fm$.
		Due to Lemmas \ref{quadratic}, \ref{pfeas}, \ref{pw} and \ref{stationarity}, the allocation vector, $\hat{x}^\fm(m^\fm)$ as $x^*$, $r_i^\fm{\bar{z}_i^{1,l}}$ as ${b_{k(i)}^{l^*}}$ (any $r_j^{\fm}{^1\bar{z}_j^l} ,\ j \in \mathcal{G}_{k(i)}^l$ could work too) and the variables $p_i^{1,l}$ and $w_i^l$ (or any $w_j^l$ for $j \in \cN^l$) as $\mu_i^{l*}$ and $\lambda^{l*}$, respectively, satisfy the  KKT conditions~\eqref{kktm}.
		Therefore, $\hat{x}^\fm(m^\fm)=x^*$ for any  $m^\fm\in \mathcal{NE}^\fm$ and hence, the allocation at all NE is unique and efficient. Also, due to Lemma \ref{exist}, we know at least one NE exists and therefore, the mechanism fully implements problem \eqref{CPm0} at its Nash equilibria.
		Furthermore, Lemma \ref{IR} proves individual rationality and weak budget balance.
	\end{IEEEproof}

	\section{Relaxing Assumption \ref{assump} on Message-Exchange Network} \label{section5}
	The primary reason of imposing Assumption \ref{assump} is that  there should be a consensus on the  prices of different agents (UTP) or groups (MMTP) using link $l$ at NE and this is not implementable by the proposed mechanism if the sub-graph of the agents using link $l$ is not connected.
	
	On the other hand, a message-exchange network that satisfies the required properties may be hard or even impossible to construct. Consider the special case of having only one link (constraint) in the UTP/MMTP optimization problem. Then, the message-exchange network should be the tree that contains all of the agents of the network. But in general there are more than one links in the UTP/MMTP optimization problem and the message-exchange network should consist of multiple connected sub-graphs (each corresponding to one constraint) and should still be a tree. Constructing such message-exchange network is hard and may be impossible.
	In this section, we propose an alternative extended mechanism so that there is no need for imposing Assumption~\ref{assump} on the message-exchange network.

	In the alternative mechanism, we extend the agents that quote message $p_i^l$, in UTP, and $w_i^l$, in MMTP, from the agents using link $l$ to a bigger group of agents as follows.  For every link $l$, consider a connected sub-graph $\mathcal{GR}^l=(\hat{\cN}^l,\mathcal{E}^l)$ consisting of all agents $i \in \cN^l$ in addition to the minimum number of other agents that do not use link $l$ and are required to make the sub-graph connected. This connected sub-graph is called link $l$'s sub-graph and we know that it exists due to the connectivity of the message graph.  For each agent $i$, the set of links $l \notin \cL_i$ which $i \in \hat{\cN}^l$ are denoted by $\hat{\cL}_i$ with $|\hat{\cL}_i|=\hat{L}_i$. The definition of $\cN^l(i)$ is modified as
	\begin{equation}
	\cN^l(i)=
	\{j:\ j \in \cN(i) \cap \hat{\cN}^l\}, \ \forall i \in \cN, l \in \cL_i \cup \hat{\cL}_i.
	\end{equation}
	In the game $\mathfrak{G}^\fu$, the extended definition of message $p_i$ is $p_i=(p_i^l, l \in \cL_i\cup \hat{\cL}_i)$, while in the game $\mathfrak{G}^\fm$, the extended definition of message $w_i$ is $w_i=(w_i^l, l  \in \cL_i\cup \hat{\cL}_i)$.
	The tax functions are also modified for $l \in \hat{\cL}_i$ according to
	\begin{equation}
	\begin{aligned}
	&\hat{t}_i^{\fu,l}(m^\fu)=\sum_{j\in\cN(i)}(n_{i,j}^l-y_j^l-  \sum_{h\in\cN(j) , h \neq i}n_{j,h}^l)^2 \\
	&+(p_i^l-\bar{p}_{-i}^l)^2
	+(p_i^l-\bar{p}_{-i}^l)\bar{p}_{-i}^l(c^l-r^\fu_if_i^{\fu,l})^2
	\end{aligned}
	\end{equation}
	
	\begin{equation}
	\begin{aligned}
	&\hat{t}_i^{\fm,l}(m^\fm)=\sum_{j\in\cN(i)}(n_{i,j}^l-y_j^l-\sum_{h\in\cN(j) , h \neq i} n_{j,h}^l)^2\\
	&+(w_i^l-\bar{w}_{-i}^l)^2+\bar{w}_{-i}^l(w_i^l-\bar{w}_{-i}^l)(c^l-r^\fm_if_i^{\fm,l})^2.
	\end{aligned}
	\end{equation}
	Intuitively, since the sub-graph of agents using each link $l$ may not be connected, we need other agents $i \notin \cN^l$ to quote $p_i^l$ messages in the game $\mathfrak{G}^\fu$ and $w_i^l$ messages in the game $\mathfrak{G}^\fm$.
	This helps the agents $j \in \cN^l$ in forming a consensus on the prices or group prices of using link $l$. This is why two terms have been added to the tax functions above that impose required conditions for the messages $p_i^l$ and $w_i^l$ in the two games.
	In both games, the tax function does not change for $l \in \cL_i$. For $l \notin \cL_i \cup \hat{\cL}_i$, the tax function is the same as the $l \notin \cL_i$ case for the original mechanism. It is straightforward to prove almost the same results (with some minor changes to cover the new messages) for these mechanisms. Therefore, these mechanisms also fully Nash implement problem \eqref{CPu} and \eqref{CPm0}, respectively, and are individually rational and weak budget balanced at NE.


	\section{Discussion and Conclusion} \label{section6}
	
	We proposed two distributed mechanisms for the networks with UTP and MMTP. As mentioned before, the mechanisms are applicable to a number of other optimization problems with linear/max constraints. The proposed mechanisms were proved to fully Nash implement the solution of problems \eqref{CPu} (UTP) and \eqref{CPm0} (MMTP). The main feature of this work is that message transmission is done locally via an underlying message-exchange network, in contrast to the standard mechanism design framework that allows message transmission throughout  the whole network.
	
	The dimensionality of agent $i$'s message  in the mechanism for UTP is $M_i=1+N(i)L+|\cI_i|+L_i$. Since for each agent $i$, the function $\phi(i)$ chooses one agent $j \in \cN(i)$, the average size of the set $\cI_i$ is $1$. Hence, the average dimensionality of an agent's  message is  $\mathbb{E}_{i\in \cN}[M_i]=2+\mathbb{E}_{i\in \cN}[N(i)]L+\mathbb{E}_{i\in \cN}[L_i]$ and by denoting $\mathbb{E}_{i\in \cN}[N(i)]$ and $\mathbb{E}_{i\in \cN}[L_i]$ by $\bar{N}$ and $\bar{L}$ respectively, the average message dimensionality of the whole network is
	\vspace{-0.2cm}
	\begin{equation*}
	\mathbb{E}[\sum_{i\in \cN}M_i]=N(2+L\bar{N}+\bar{L}).
	\end{equation*}
	Clearly, the dimensionality of message space grows linearly with $N$.
	
	The dimensionality of agent $i$'s message  in the mechanism for MMTP is $M_i=1+4L_i+N(i)L+|\cI_i|+2\sum_{j \in \cI_i}L_j+2|\mathcal{C}_i|$. Similar to UTP, the average size of the set $\cI_i$ is $1$. Further, the average value of $\sum_{j \in \cI_i}L_j$ is $\frac{\sum_{i \in \cN}L_i}{N}$. Also, we know that for each link $l$ and group $k \in \mathcal{K}^l$, there is one agent denoted by $c(k,l)$ and hence, the average size of $|\mathcal{C}_i|$ is $\frac{\sum_{l \in \cL}K^l}{N}$.  Consequently, the average size of  the whole network's message is
	\vspace{-0.2cm}
	\begin{align*}
	\mathbb{E}[\sum_{i\in \cN}M_i]=N(2+4\bar{L}+\bar{N}L+2\frac{\sum_{i \in \cN}L_i}{N}+2\frac{\sum_{l \in \cL}K^l}{N}),
	\end{align*}
	which, similar to UTP, grows linearly with the number of agents in the network, $N$.

	For the alternative mechanism presented in Section~\ref{section5}, the term $N\mathbb{E}_{i \in \cN}(\hat{L}_i)$ should be added to the average of the message dimensionality of the whole network. This is due to the extra messages that agents have to quote in the alternative mechanisms to preserve the connectivity of the message passing.
	
	In terms of message dimensionality, the mechanisms proposed in this paper are more efficient than the distributed mechanism proposed in \cite{SiAn17d} which has a message dimensionality that grows with $N^2$. This gain in dimensionality may be a consequence of the fact that the proposed mechanism in \cite{SiAn17d} has additional learning guarantees that our proposed mechanism does not possess.

	We would like to emphasize that although the proposed mechanisms are proven to have efficient Nash equilibria, in the current work we do not propose any message exchange algorithm that is guaranteed to converge to these equilibria. In general, it is not even clear if such algorithm exists.
	We have recently proposed one algorithm that is guaranteed to converge to the NE for our mechanisms with some small modifications to the model in~\cite {ARXIV-VERSION-J}. However,  the study of more general convergence algorithms is still open and there is no strategy-proof dynamic algorithm that guarantees convergence.
	One possible future research direction is redesigning these mechanisms to possess other features such as learning guarantees and convergence to NE.
	Such features would enable the mechanisms to converge to their NE in a dynamic learning process over a large set of possible dynamics followed by the agents~\cite{healy2012designing,SiAn17d}.
	In addition, the study of the tradeoff between message dimensionality and convergence guarantees is an interesting open problem.

	\bibliographystyle{IEEEtran}

	\vspace*{-0.8cm}

\begin{IEEEbiography}[{\includegraphics[width=1in,height=1.25in,clip,keepaspectratio]{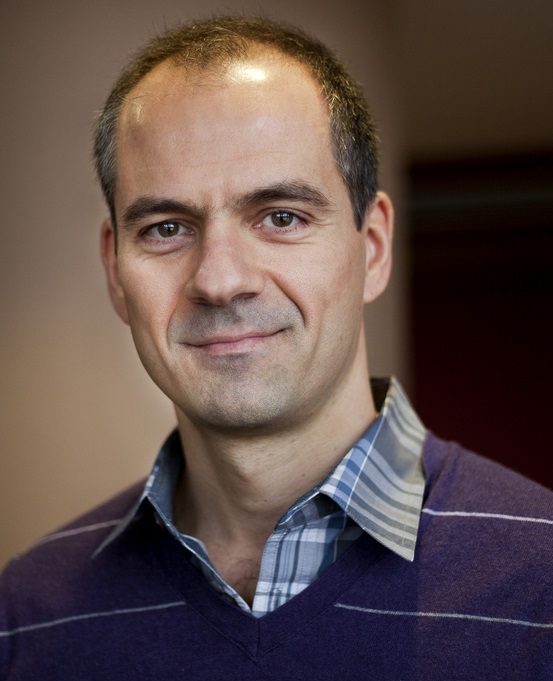}}]{Achilleas Anastasopoulos}
(S'97-M'99-SM'13) was born in Athens, Greece in 1971. He received the Diploma
in Electrical Engineering from the National Technical University of
Athens, Greece in 1993, and the M.S.\ and Ph.D.\ degrees in Electrical
Engineering from University of Southern California in 1994 and 1999,
respectively. He is currently an Associate Professor at the University of
Michigan, Ann Arbor, Department of Electrical Engineering and Computer
Science.

His research interests lie in the general area of communication and information theory,
with emphasis in channel coding and multi-user channels;
control theory with emphasis in decentralized stochastic control and its
connections to communications and information theoretic problems;
analysis of dynamic games and mechanism design for resource allocation on networked systems.

He is the co-author
of the book \emph{Iterative Detection: Adaptivity, Complexity Reduction,
and Applications,} (Reading, MA: Kluwer Academic, 2001).

Dr.\ Anastasopoulos is the recipient of the ``Myronis Fellowship'' in
1996 from the Graduate School at the University of Southern California,
the NSF CAREER Award in 2004,
and was a co-author for the paper that received the best student paper award in ISIT 2009.
He served as a technical program committee
member for ICC , Globecom, VTC, and ISIT conferences, and on the editorial
board of the IEEE TRANSACTIONS ON COMMUNICATIONS.
\end{IEEEbiography}

	\vspace*{-0.9cm}
	\begin{IEEEbiography}[{\includegraphics[width=1in,height=1.25in,clip,keepaspectratio]{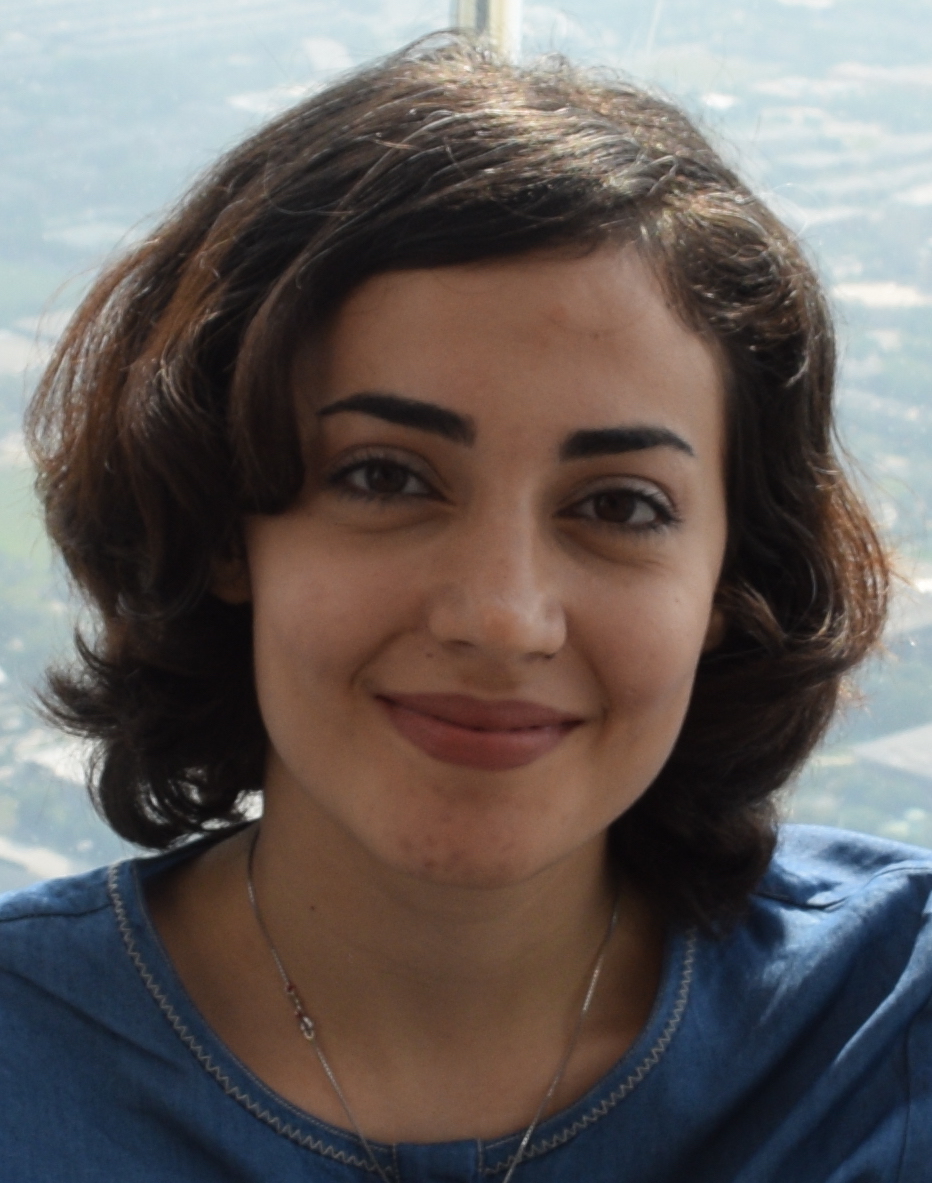}}]{Nasimeh Heydaribeni}
was born in Esfahan, Iran in 1993. She received B.S. and M.S. degrees in Electrical Engineering in Sharif University of Technology, Tehran, Iran, in 2015 and 2017, respectively. She is currently persuing her Ph.D. degree in Electrical Engineering and Computer Science in University of Michigan, Ann Arbor. Her research interests are mechanism design for resource allocation in networks and game theory and its applications in networked systems.
\end{IEEEbiography}
\vfill


	\clearpage
	
	\section*{Appendix}\label{appendix}
	\subsection{Proof of Lemma \ref{con}}
	Since functions $\hat{u}^\fu_i(m_i^\fu,m_{-i}^\fu)$ and $\hat{u}^\fm_i(m_i^\fm,m_{-i}^\fm)$ are twice differentiable w.r.t $m_i^\fu$ and $m_i^\fm$, respectively, we can prove strict concavity
	by showing that their Hessian matrices, $\mathrm{H}^\fu$ and $\mathrm{H}^\fm$,  w.r.t. $m_i^\fu$ and $m_i^\fm$, respectively,  are negative definite.
	The cross derivatives of  $\hat{u}^\fu_i(m_i^\fu,m_{-i}^\fu)$ and $\hat{u}^\fm_i(m_i^\fm,m_{-i}^\fm)$ w.r.t. different components of $m_i^\fu$ and $m_i^\fm$, which are the non-diagonal elements of $\mathrm{H}^\fu$ and $\mathrm{H}^\fm$, respectively, are zero. Hence, we consider the diagonal elements and show that they are all negative.
	The second partial derivative of  $\hat{u}^\fu_i(m^\fu)$ w.r.t. all elements of messages $n_i$, $q_i$, $p_i$ is equal to $-2$. Also, the second partial derivative of  $\hat{u}^\fm_i(m^\fm)$ w.r.t. all elements of messages $n_i$, $q_i$, $p_i$, $w_i$, $z_i$, $a_i$ and $\yy_i$ is equal to $-2$. The only message element left is $y_i$.
	The second partial derivative of $\hat{u}^\fu_i(m^\fu)$ w.r.t. $y_i$ is ${\partial^2 \hat{u}^\fu_i(m^\fu)}/{(\partial y_i)^2}=(r_i^\fu)^2{\partial^2 v_i(\hat{x}^\fu_i)}/{(\partial \hat{x}^\fu_i)^2}$. Since $v_i(\hat{x}^\fu_i)$ is strictly concave w.r.t. $\hat{x}^\fu_i$,  ${\partial^2 \hat{u}^\fu_i(m^\fu)}/{(\partial y_i)^2}<0$.
	Similarly, the second partial derivative of $\hat{u}^\fm_i(m^\fm)$ w.r.t. $y_i$ is ${\partial^2 \hat{u}^\fm_i(m^\fm)}/{(\partial y_i)^2}=(r_i^\fm)^2{\partial^2 v_i(\hat{x}^\fm_i)}/{(\partial \hat{x}^\fm_i)^2}<0$.
	Note that $r_i^\fu$ and $r_i^\fm$ don't consist of any of agent $i$'s messages and so they are constant factors.
	
	Therefore, matrices $\mathrm{H}^\fu$ and $\mathrm{H}^\fm$ are negative definite because all of their diagonal elements are negative and non-diagonal elements are zero.

	\subsection{Proof of Lemma \ref{qij1}}
	At NE, every agent is best responding to other agents' messages.
	Each of the results in this lemma corresponds to one of agent $i$'s messages and its best response to other agents messages. Therefore,  all of the results can be directly derived by setting each of their corresponding element of gradient to zero.
	For all $i \in \cN$ we have
	\begin{equation*}
	\frac{\partial\hat{u}^\fu_i(m_i^\fu,m_{-i}^\fu)}{\partial q_{i,j}}=0
	\Rightarrow
	q_{i,j}=y_j, \
	\forall j \in \cI_i.
	\end{equation*}
	Since $y_j\geq 0$ the above equation can always hold.
	
	\begin{align}
	&\frac{\partial\hat{u}^\fu_i(m_i^\fu,m_{-i}^\fu)}{\partial n_{i,j}^l} =0
	\Rightarrow \nonumber \\
	&n_{i,j}^l =y_j^l+\sum_{h\in\cN(j) , h \neq i} n_{j,h}^l \ \ \forall  j \in \cN(i), l \in \cL.
	\label{nproof}
	\end{align}		
	Using a similar argument as the one used in \cite[p. 131]{Si17} all above equations can be combined to show that
	\begin{equation}
	n_{i,j}^l=\sum_{h\in\cN , n(i,h)=j}y_h^l,
	\label{ncon}
	\end{equation}
	and equivalently,
	\begin{equation*}
	\sum_{j \in \cN(i)}n_{i,j}^l=\sum_{h \in \cN, h\neq i} y_h^l.
	\end{equation*}
	
	In order to verify this conclusion, we mention that the message graph is a tree, and hence we can form an induction on the level of nodes from the leaf nodes. If $j \in \cN(i)$ is a leaf node, there is no message components $n_{j,h}^l$ for $h\in\cN(j) , h \neq i$. because the only neighbor of $j$ is $i$. As a result, $n_{i,j}^l=y_j^l$ and therefore, the induction basis holds.
	Suppose for all $j \in \cN(i)$,  $n_{j,h}^l=\sum_{k\in\cN , n(j,k)=h}y_k^l, \ \forall h \in \cN(j)$. Substituting this to~\eqref{nproof} we have
	\begin{align}
	n_{i,j}^l =y_j^l+\sum_{\substack{h\in\cN(j) \\ h \neq i}}\sum_{\substack{k\in\cN \\ n(j,k)=h}}y_k^l \ \ \forall  j \in \cN(i), l \in \cL.
	\label{nproof2}
	\end{align}
	
	We need to check whether the set of nodes covered in the summations above is equal to the set of nodes in the summation of~\eqref{ncon}. In~\eqref{nproof2}, we are summing over all nodes $k$ that can be reached to node $j$ by nodes  $h \in \cN(j), \ h \neq i$. Since $j \in \cN(i)$ and there is only one path between any two nodes in the graph, we conclude that all of these nodes reach $i$ through $j$ and therefore, $n(i,k)=j$. This means that the summations in~\eqref{ncon} and~\eqref{nproof2} include the same set of nodes and the result is proved.
	\subsection{Proof of Lemma \ref{quadratic}}
	According to the explanations at the beginning of the proof of Lemma \ref{qij1}, by setting each of the corresponding element of gradients of $\hat{u}^\fm_i(m_i^\fm,m_{-i}^\fm)$ to zero for all $i \in \cN$ we have
	
	\begin{align*}
	&\frac{\partial\hat{u}^\fm_i(m_i^\fm,m_{-i}^\fm)}{\partial q_{i,j}}=0
	\Rightarrow
	q_{i,j}=y_j, \
	\forall  j \in \cI_i\\
	&\frac{\partial\hat{u}^\fm_i(m_i^\fm,m_{-i}^\fm)}{\partial(\yy_i^l)}=0 \Rightarrow
	\yy_i^l=\frac{q_{\phi(i),i}\textbf{1}_{\{q_{\phi(i),i}\}}( \bar{z}_i^{1,l})}{\bar{z}_i^{2,l}},
	\ \forall  l \in \cL_i \\
	&\frac{\partial\hat{u}^\fm_i(m_i^\fm,m_{-i}^\fm)}{\partial (p_{i,j}^{2,l})}=0
	\Rightarrow
	{p_{i,j}^{2,l}}=p_j^{1,l},
	\ \forall  j \in {I}_i,  l \in \cL_j\\
	&\frac{\partial\hat{u}^\fm_i(m_i^\fm,m_{-i}^\fm)}{\partial w_i^l}=0
	\Rightarrow
	w_i^l=w_{c(k(i),l)}^l,
	\  \forall  l \in \cL_i , l \notin \mathcal{C}_i\\
	&\frac{\partial\hat{u}^\fm_i(m_i^\fm,m_{-i}^\fm)}{\partial w_i^l}=0
	\Rightarrow \\
	&\quad w_i^l={p_{\phi(i),i}^{2,l}}+\sum_{j \in \mathcal{G}_{k(i)}^l, j \neq i}{p_j^{1,l}},\ \  \forall  l \in \cL_i, l \in \mathcal{C}_i\\
	&\frac{\partial\hat{u}^\fm_i(m_i^\fm,m_{-i}^\fm)}{\partial n_{i,j}^l}=0
	\Rightarrow
	n_{i,j}^l=y_j^l+\sum_{h\in\cN(j) , h \neq i}n_{j,h}^l, \\
	& \hspace*{5cm} \forall  j \in \cN(i), l \in \cL. \nonumber
	\end{align*}
	Using a similar argument as the one used in \cite[p. 131]{Si17} we prove that
	\begin{equation*}
	n_{i,j}^l=\sum_{h\in\cN , n(i,h)=j}y_h^l,
	\end{equation*}
	and consequently,
	\begin{equation*}
	\sum_{j \in \cN(i)}n_{i,j}^l=\sum_{h\in\cN, h \neq i}y_h^l.
	\end{equation*}
	The remaining results are related to message elements $z_i$ and ${^2a}_i$.
	\begin{align*}
	&\frac{\partial\hat{u}^\fm_i(m_i^\fm,m_{-i}^\fm)}{\partial ({z_i^{1,l}})}=0
	\Rightarrow
	{z_i^{1,l}}= \bar{z}_i^{1,l},
	\quad \forall  l \in \cL_i , l \in \mathcal{C}_i\\
	&\frac{\partial\hat{u}^\fm_i(m_i^\fm,m_{-i}^\fm)}{\partial (z_i^{2,l})}=0
	\Rightarrow
	{z_i^{2,l}}=\bar{z}_i^{2,l},
	\quad  \forall  l \in \cL_i , l \in \mathcal{C}_i\\
	&\frac{\partial\hat{u}_i^\fm(m_i^\fm,m_{-i}^\fm)}{\partial ({a_{i,j}^{2,l}})}=0
	\Rightarrow
	{a_{i,j}^{2,l}}=a_j^{1,l},  \quad  \forall  j \in \cI_i, l \in \mathcal{L}_j.
	\end{align*}
	\subsection{Proof of Lemma \ref{pfeas}}
	According to Lemmas \ref{qij1} and \ref{quadratic}, the following relation holds at NE,
	\begin{equation*}
	f_i^{\fu,l}=\sum_{j \in \cN}y_j^l,
	\end{equation*}
	and similarly,
	\begin{equation*}
	f_i^{\fm,l}=\sum_{j \in \cN}y_j^l.
	\end{equation*}
	This implies that all of the agents $i \in \cN$ have the same $f_i^{\fu,l}$ and $f_i^{\fm,l}$ and consequently, have the same $r_i^\fu$ and $r_i^\fm$ at NE of the games $\mathfrak{G}^\fu$ and $\mathfrak{G}^\fm$, respectively. Further, due to Lemma \ref{quadratic}, for the game $\mathfrak{G}^\fm$ we can write for every $j \in \mathcal{G}_k^l$
	\begin{equation*}
	y_j^l=\left\{
	\begin{array}{cc}
	\frac{y_j}{n\_\textrm{max}_k^l} & \ \textrm{if} \ l \in \cL_j, \ y_j=\max_{i \in \mathcal{G}_{k}^l}\{y_i\}\\
	0 & \ \textrm{oth.},
	\end{array}\right.
	\end{equation*}
	where $n\_\textrm{max}_k^l$ is the number of agents $j \in \mathcal{G}_k^l$ with $y_j=\max_{i \in \mathcal{G}_{k}^l}\{y_i\}$.
	Consequently, $\forall l \in \cL$ we can write
	\begin{equation*}
	\begin{aligned}
	&\sum_{k\in \mathcal{K}^l} \max_{i \in \mathcal{G}_k^l} \{\hat{x}^\fm_i\}=\sum_{k\in \mathcal{K}^l} \max_{i \in \mathcal{G}_k^l}\{r_i^\fm y_i\}   &\\ & \leq \frac{c^l}{\sum_{j \in \cN}y_j^l}\sum_{k\in \mathcal{K}^l} \max_{i \in \mathcal{G}_k^l}\{y_i\}
	=\frac{c^l}{\sum_{j \in \cN}y_j^l}\sum_{i \in \cN}y_i^l=c^l,
	\end{aligned}
	\end{equation*}
	which proves that the allocation $\hat{x}^\fm$ is feasible at NE.	
	
	By using similar steps, we can show the feasibility of allocation of the game $\mathfrak{G}^\fu$ at NE.


	\subsection{Proof of Lemma \ref{pil1}}
	In order to prove the first result, we first derive the following
	\begin{equation*}
	p_i^l = \bar{p}_{-i}^l  \quad \forall i \in \cN , l \in \cL_i.
	\end{equation*}
	Suppose the above equation does not hold, i.e.,
	\begin{equation*}
	\exists i \in \cN , l \in \cL_i : \quad	p_i^l \neq \bar{p}_{-i}^l.
	\end{equation*}
	Then there exists an agent $j \in \cN^l: p_j^l>\bar{p}_{-j}^l$ (as an example, we could consider the agent $j$ with the highest $p_j^l$ over all of the agents and if we have multiple choices, at least one of them will work). We can show that agent $j$ has a profitable deviation to $p_j^{l'}=\bar{p}_{-j}^l=p_j^l-\epsilon$. Indeed, we can write
	\begin{equation*}
	\begin{aligned}
	&\hat{u}_j^\fu(.,p_j^{l'})
	-\hat{u}^\fu_j(.,p_j^l)
	=\epsilon^2+\epsilon\bar{p}_{-j}^l(c^l-r_j^\fu f_j^{\fu,l})^2 \\&=\epsilon(\underbrace{\epsilon}_{>0}+\underbrace{\bar{p}_{-j}^l(c^l-r_j^\fu f_j^{\fu,l})^2}_{\geq 0})>0,
	\end{aligned}
	\end{equation*}
	therefore, we must have $p_i^l=\bar{p}_{-i}^l$.
	
	As a result of this equality and because of Assumption~\ref{assump}, it is obvious that $p_i^l=p_j^l, \ \forall i,j \in \cN^l$ and we denote this common price by $p^l$.
	
	For the second result, 	we set the derivative of the utility function w.r.t. $p_i^l$ to zero,
	\begin{equation*}
	\begin{aligned}
	&\frac{\partial\hat{u}_i^\fu(m_i^\fu,m_{-i}^\fu)}{\partial p_i^l}=0 \Rightarrow \underbrace{2(p_i^l-\bar{p}_{-i}^l)}_{=0 , \textrm{Due to~\eqref{eq:eq_prices}}}+\bar{p}_{-i}^l(c^l-r_i^\fu f_i^{\fu,l})^2 =0 \\
	&\Rightarrow
	p^l(c^l-r_i^\fu f_i^{\fu,l})^2=0
	\Rightarrow p^l(c^l-\sum_{i \in \cN^l}\hat{x}_i^\fu)=0.
	\end{aligned}
	\end{equation*}
	%

	\subsection{Proof of Lemma \ref{pw}}
	We first prove result \eqref{eqw}. This result is equivalent with $\hat{w}_i^l=\hat{w}_j^l, \ \forall i,j \in \cN^l$.
	Assume $ \exists i,j \in \mathcal{N}^l, \
	\hat{w}_i^l \neq \hat{w}_j^l$. Since $w_i^l=\hat{w}_i^l$ at NE and due to Assumption~\ref{assump}, there exists an agent $h \in \mathcal{N}^l$ for which $\hat{w}_h^l>\bar{w}_{-h}^l$ or equivalently, $\hat{w}_h^l=\bar{w}_{-h}^l+\epsilon$ for some $\epsilon>0$. We will show that agent $h$ has a profitable deviation by decreasing his message $a_h^{1,l}$ to $a_h^{1,l'}=a_h^{1,l}-\epsilon'>0$ for some $0<\epsilon'<\epsilon$. Consequently, ${\hat{w}_h^{l'}}=\hat{w}_h^l-\epsilon'=\bar{w}_{-h}^l+\epsilon-\epsilon'=\bar{w}_{-h}^l+\epsilon''$. We can write
	\begin{align*}
	&\hat{u}_h^\fm(.,a_h^{1,l'})
	-\hat{u}_h^\fm(.,a_h^{1,l})=\\&
	-\epsilon''^2-\bar{w}_{-h}^l\epsilon''(c^l-r_h^\fm f_h^{\fm,l})^2+\epsilon^2+\bar{w}_{-h}^l\epsilon(c^l-r_h^\fm f_h^{\fm,l})^2\\&=\underbrace{\epsilon^2-\epsilon''^2}_{>0}+
	\underbrace{\bar{w}_{-h}^l(\epsilon-\epsilon'')(c^l-r_h^\fm f_h^{\fm,l})^2}_{\geq 0}>0,
	\end{align*}
	and we conclude that at any NE, $\hat{w}_i^l=\hat{w}_j^l, \ \forall l \in \cL, i,j \in \mathcal{N}^l$. Therefore, we can denote this common value for each link $l$ by $w^l$ and we arrive at the result $\hat{w}_i^l=w^l, \ \forall i \in \cN, \ l \in \cL_i$.
	
	Now we prove result \eqref{comw}. Suppose $\exists i \in \mathcal{N}, l \in \mathcal{L}_i$ so that $w^l(c^l-r_i^\fm \sum_{i \in \cN}y_i^l) \neq 0$. This implies $\bar{w}_{-i}^l(c^l-r_i^\fm f_i^{\fm,l})^2>0$. We show that agent $i$ benefits from deviating to $a_i^{1,l'}=a_i^{1,l}-\epsilon>0$, for some $\epsilon>0$. According to the first result of this lemma, $\hat{w}_i^l=\bar{w}_{-i}^l$ and we have
	\begin{align*}
	&\hat{u}_i^\fm(.,a_i^{1,l'})
	-\hat{u}_i^\fm(.,a_i^{1,l})=
	-\epsilon^2+\bar{w}_{-i}^l\epsilon(c^l-r_i^\fm f_i^{\fm,l})^2
	\\&=\epsilon(-\epsilon+\underbrace{\bar{w}_{-i}^l(c^l-r_i^\fm f_i^{\fm,l})^2}_{> 0,\  \textrm{Due to assumption}})=\epsilon(-\epsilon+\alpha)>0, \  \textrm{for} \ \epsilon<\alpha.
	\end{align*}
	Since $\alpha>0$, profitable deviation by a positive $\epsilon$ is possible and the result is proved.
	
	Proving result \eqref{comp} is similar to result \eqref{comw}. Assume $\exists i \in \mathcal{N}, l \in \mathcal{L}_i$ so that ${p_i^{1,l}}(y_i-\bar{z}_i^{1,l}) \neq 0$. This implies that $p_{\phi(i),i}^{2,l}(\bar{z}_i^{1,l}-q_{\phi(i),i})^2 > 0$ and  ${p_i^{1,l}}>0$. We prove agent $i$ has a profitable deviation to  ${p_i^{1,l'}}={p_i^{1,l}}-\epsilon>0$, for some $\epsilon>0$.
	Indeed,
	\begin{align*}
	&\hat{u}_i^\fm(.,p_i^{1,l'})
	-\hat{u}_i^\fm(.,p_i^{1,l})=
	-\epsilon^2+p_{\phi(i),i}^{2,l}\epsilon(\bar{z}_i^{1,l}-q_{\phi(i),i})^2\\
	&=\epsilon(-\epsilon+\underbrace{p_{\phi(i),i}^{2,l}(\bar{z}_i^{1,l}-q_{\phi(i),i})^2}_{> 0, \ \textrm{Due to assumption}})=\epsilon(-\epsilon+\alpha)>0,  \textrm{for} \ \epsilon<\alpha.
	\end{align*}
	Since $\alpha>0$, agent $i$ can profit by deviating with a positive $\epsilon$ and the result is proved.

	\subsection{Proof of Lemma \ref{stationarity1}}
	If $\hat{x}_i^\fu(m^\fu)>0$, then $y_i>0$ and hence, the partial derivative of $\hat{u}_i^\fu(m_i^\fu,m_{-i}^\fu)$ w.r.t. $y_i$ must be zero at NE. Therefore,
	\begin{equation*}
	\begin{aligned}
	&\frac{\partial\hat{u}_i^\fu(m_i^\fu,m_{-i}^\fu)}{\partial y_i}=0
	\Rightarrow (\frac{\partial\hat{u}_i^\fu(m_i^\fu,m_{-i}^\fu)}{\partial \hat{x}_i^\fu(m^\fu)})\frac{d\hat{x}_i^\fu(m^\fu)}{dy_i}=0 \\
	&=
	(v_i^\prime(\hat{x}_i^\fu(m^\fu))-\sum_{l\in\cL_i}p^l)r_i^\fu
	\Rightarrow v_i^\prime(\hat{x}_i^\fu(m^\fu))=\sum_{l\in\cL_i}p^l,
	\end{aligned}
	\end{equation*}
	and	if $\hat{x}_i^\fu(m^\fu)=0$, then $y_i=0$ and therefore, the partial derivative of $\hat{u}_i^\fu(m_i^\fu,m_{-i}^\fu)$ w.r.t. $y_i$ must not be positive at NE. Hence,
	\begin{equation*}
	\begin{aligned}
	&\frac{\partial\hat{u}_i^\fu(m_i^\fu,m_{-i}^\fu)}{\partial y_i}\leq 0
	\Rightarrow (\frac{\partial\hat{u}_i^\fu(m_i^\fu,m_{-i}^\fu)}{\partial \hat{x}_i^\fu(m^\fu)})\frac{d\hat{x}_i^\fu(m^\fu)}{dy_i} \\
	&
	=(v_i^\prime(\hat{x}_i^\fu(m^\fu))-\sum_{l\in\cL_i}p^l)r_i^\fu\leq0
	\Rightarrow v_i^\prime(\hat{x}_i^\fu(m^\fu))\leq \sum_{l\in\cL_i}p^l.
	\end{aligned}
	\end{equation*}

	\subsection{Proof of Lemma \ref{stationarity}}
	Similar to Lemma \ref{stationarity1}, if $\hat{x}_i^\fm(m^\fm)>0$,
	\begin{equation*}
	\begin{aligned}
	&\frac{\partial\hat{u}_i^\fm(m_i^\fm,m_{-i}^\fm)}{\partial y_i}=0
	\Rightarrow (\frac{\partial\hat{u}_i^\fm(m_i^\fm,m_{-i}^\fm)}{\partial \hat{x}_i^\fm(m^\fm)})\frac{d\hat{x}_i^\fm(m^\fm)}{dy_i}=0\\ &\Rightarrow
	(v_i^\prime(\hat{x}_i^\fm(m^\fm))-\sum_{l\in\cL_i}p_{\phi(i),i}^{2,l})r_i^\fm=0\\
	&\Rightarrow v_i^\prime(\hat{x}_i^\fm(m^\fm))=\sum_{l\in\cL_i}p_i^{1,l}.
	\end{aligned}
	\end{equation*}
	Note that $r_i^\fm>0$.
	If $\hat{x}_i^\fm(m^\fm)=0$,
	\begin{equation*}
	\begin{aligned}
	&\frac{\partial\hat{u}_i^\fm(m_i^\fm,m_{-i}^\fm)}{\partial y_i}\leq0
	\Rightarrow (\frac{\partial\hat{u}_i^\fm(m_i^\fm,m_{-i}^\fm)}{\partial \hat{x}_i^\fm(m^\fm)})\frac{d\hat{x}_i^\fm(m^\fm)}{dy_i} \leq0\\ &\Rightarrow
	(v_i^\prime(\hat{x}_i^\fm(m^\fm))-\sum_{l\in\cL_i}p_{\phi(i),i}^{2,l})r_i^\fm\leq0\\
	&\Rightarrow v_i^\prime(\hat{x}_i^\fm(m^\fm))\leq\sum_{l\in\cL_i}p_i^{1,l}.
	\end{aligned}
	\end{equation*}
	
	\subsection{Proof of Lemma \ref{exist}}
	To prove the existence of a NE, we show that a suggested valid message is a NE. For each of the games $\mathfrak{G}^\fu$ and $\mathfrak{G}^\fm$, the suggested message is generated based on the solution of problems \eqref{CPu} and \eqref{CPm0}, respectively, which we know exist and is unique.
	We notice that because of the monotonicity of valuation functions, the solution of problems \eqref{CPu} and \eqref{CPm0} always lies in the Pareto optimal region of the feasible set which, in our case, is the upper boundary of feasible set in both UTP and MMTP.
	First consider the game $\mathfrak{G}^\fu$.
	Suppose $(x^*,\lambda^*)$ is the solution of problem \eqref{CPu}. We generate $m^{\fu}$ as follows. First assume $m^{\fu}$ satisfies all of the constraints in Lemma \ref{qij1}. Further, $y$ is set to be any scaled version of $x^*$ and since $x^*$ is on the boundary of feasible region, $\hat{x}^\fu(m^{\fu})=x^*$. In addition, $p_i^l$ is set to be equal to $\lambda^{l*}$ and this is valid since $\lambda^{l*}\geq0$.
	Hence, Lemma \ref{pil1} is satisfied for $m^{\fu}$. Also, due to stationarity condition, Lemma \ref{stationarity1} is also satisfied for $m^{\fu}$. Overall, since Lemmas  \ref{qij1}, \ref{pil1} and \ref{stationarity1} are satisfied, we know that the elements of the gradient vector of agent $i$'s utility function w.r.t. $m_i^{\fu}$ is either zero (positive messages) or not positive (zero messages) which implies that each agent is best responding to other agents' messages and therefore, $m^{\fu}$ is a NE of the game $\mathfrak{G}^\fu$.
	
	Similar steps are taken for the proof of existence of NE for the game $\mathfrak{G}^\fm$.
	Let $(x^*,b^*,\lambda^*,\mu^*)$ be the solution of problem \eqref{CPm}.
	We generate $m^{\fm}$ as following.
	First assume $m^{\fm}$ satisfies all of the constraints in Lemma \ref{quadratic}.
	Further, $y$ is set to be any scaled version of $x^*$ and since $x^*$ is on the boundary of feasible region, $\hat{x}^\fm(m^\fm)=x^*$. In addition, ${p_i^{1,l}}$ is set to be equal to $\mu_i^{l*}$ and this is valid because $\mu_i^{l*}\geq0$.
	Then, $w^l=\sum_{j \in \mathcal{G}_{k}^l}{p_j^{1,l}}=\sum_{j \in \mathcal{G}_{k}^l}\mu_j^{l*}=\lambda^{l*}$.
	Also $r_i^\fm \bar{z}_i^{1,l}=\max_{j \in \mathcal{G}_{k(i)}^l}\{r_i^\fm y_j\}=\max_{j \in \mathcal{G}_{k(i)}^l}\{\hat{x}_i^\fm\}=\max_{j \in \mathcal{G}_{k(i)}^l}\{x_i^*\}=b_{k(i)}^{l*}$.
	Hence Lemma \ref{pw} is satisfied for $m^{\fm}$.
	Also, due to stationarity condition, Lemma \ref{stationarity} is satisfied for $m^{\fm}$.
	Overall, since Lemmas  \ref{quadratic}, \ref{pw} and \ref{stationarity} are satisfied, we know that the elements of the gradient vector of utility function of each agent $i$ w.r.t. $m_i^{\fm}$ is either zero (positive messages) or not positive (zero messages) which implies that each agent is best responding to other agents' messages and therefore, $m^{\fm}$ is a NE.
	
	Notice that the dual variables in the solution of optimization problems \eqref{CPu} and \eqref{CPm} are not unique, eventhough the primal solution ($x$) is unique. For each game and each value of dual variables there is a suggested message that is a Nash equilibrium for that game. Further, the $y$ messages at Nash equilibria of these games have infinitely many options as it was mentioned in its construction. This means that the Nash equilibria of these games are not unique and in fact there are infinitely many Nash equilibria for these games.
	
	\subsection{Proof of Lemma \ref{IR}}
	First consider the weak budget balance equations. At NE, we can write $\hat{t}^\fu_i=\hat{x}^\fu_i(m^\fu)\sum_{l \in \cL_i}p^l$ and hence $\sum_{i \in \cN}\hat{t}^\fu_i\geq 0$.
	Similarly, $\hat{t}^\fm_i=\hat{x}^\fm_i(m^\fm)\sum_{l \in \cL_i}{p_i^{1,l}}$ and hence, $\sum_{i \in \cN}\hat{t}^\fm_i\geq 0$ and both mechanisms are weak budget balanced.
	
	Next, consider the individual rationality part for UTP mechanism (the MMTP version is almost identical).
	For $\hat{x}^\fu_i(m^\fu)=0$, the result is obvious.
	For $\hat{x}^\fu_i(m^\fu)>0$, we define the function $u_i$ as
	\begin{align*}
	u_i(x)&=v_i(x)-x\sum_{l \in \cL_i}p^l.
	\end{align*}
	Since $u_i(x)$ is concave w.r.t. $x$ and $u_i'(\hat{x}^\fu_i(m^\fu))=0$, then $u_i'(y)\geq0$ for $0\leq y\leq\hat{x}^\fu_i(m^\fu)$, we can conclude $u_i(y)\geq u_i(0)$
	and since $u_i(0)=v_i(0)$ and $u_i(\hat{x}^\fu_i(m^\fu))=v_i(\hat{x}^\fu_i(m^\fu))-\hat{t}^\fu_i(m^\fu)$, it follows that 	$v_i(\hat{x}^\fu_i(m^\fu))-\hat{t}^\fu_i(m^\fu)\geq v_i(0)$ and the result is proved.

%

\end{document}